\pdfoutput=1
\documentclass[
    superscriptaddress,
    amsmath, amssymb,
    reprint,
    aps,
    prx
]{revtex4-2}

\usepackage{physics}
\usepackage{graphicx}
\usepackage{float}
\usepackage{dcolumn} 
\usepackage{hyperref}
\usepackage{hyperref}
\hypersetup{
    colorlinks = true,
    allcolors = {blue}
}
\usepackage[capitalise]{cleveref}
\usepackage[normalem]{ulem}
\usepackage{tabularx}

\usepackage{xcolor}

\begin{document}

\title{Floquet Mode Traveling Wave Parametric Amplifiers}

\author{Kaidong Peng}
\affiliation{Department of Electrical Engineering and Computer Science, Massachusetts Institute of Technology, Cambridge, MA 02139, USA}
\affiliation{Research Laboratory of Electronics, Massachusetts Institute of Technology, Cambridge, MA 02139, USA}
\author{Mahdi Naghiloo}
\affiliation{Research Laboratory of Electronics, Massachusetts Institute of Technology, Cambridge, MA 02139, USA}
\author{Jennifer Wang}

\affiliation{Department of Electrical Engineering and Computer Science, Massachusetts Institute of Technology, Cambridge, MA 02139, USA}
\affiliation{Research Laboratory of Electronics, Massachusetts Institute of Technology, Cambridge, MA 02139, USA}

\author{Gregory D. Cunningham}

\affiliation{Research Laboratory of Electronics, Massachusetts Institute of Technology, Cambridge, MA 02139, USA}

\affiliation{Harvard John A. Paulson School of Engineering and Applied Sciences, Harvard University, Cambridge, MA 02138, USA}

\author{Yufeng Ye}
\author{Kevin P. O'Brien}
\email[Correspondence email address:]{\ kpobrien@mit.edu}

\affiliation{Department of Electrical Engineering and Computer Science, Massachusetts Institute of Technology, Cambridge, MA 02139, USA}
\affiliation{Research Laboratory of Electronics, Massachusetts Institute of Technology, Cambridge, MA 02139, USA}

\date{\today}

\begin{abstract}
~\\~\\
Simultaneous ideal quantum measurements of multiple single-photon-level signals would advance applications in quantum information processing, metrology, and astronomy, but require the first amplifier to be simultaneously broadband, quantum limited, and directional. However, conventional traveling-wave parametric amplifiers support broadband amplification at the cost of increased added noise and are not genuinely directional due to non-negligible nonlinear backward wave generation. In this work, we introduce a new class of amplifiers which encode the information in the Floquet modes of the system. Such Floquet mode amplifiers prevent information leakage and overcome the trade-off between quantum efficiency (QE) and bandwidth. Crucially, Floquet mode amplifiers strongly suppress the nonlinear forward-backward wave coupling and are therefore genuinely directional and readily integrable with qubits, clearing another major obstacle towards broadband ideal quantum measurements. Furthermore, Floquet mode amplifiers are insensitive to out-of-band impedance mismatch, which otherwise may lead to gain ripples, parametric oscillations, and instability in conventional traveling-wave parametric amplifiers. Finally, we show that a Floquet mode Josephson traveling-wave parametric amplifier implementation can simultaneously achieve $>\!20\,$dB gain and a QE of $\eta/\eta_{\mathrm{ideal}}\!> 99.9\%$ of the quantum limit over more than an octave of bandwidth. The proposed Floquet scheme is also widely applicable to other platforms, such as kinetic inductance traveling-wave amplifiers and optical parametric amplifiers.\\~\\
\end{abstract}

\maketitle

Faithful amplification and detection of weak signals are of central importance to various research areas in fundamental and applied sciences, ranging from the study of celestial objects in radio astronomy and metrology \cite{day_a_2003, pospieszalski_extremely_2005}, dark-matter detection in cosmology \cite{asztalos_SQUID_2010, dixit_searching_2021,bartram_dark_2021}, and exploration of novel light-matter interactions in atomic physics \cite{forn_ultrastrong_2019} to superconducting \cite{mallet_single_2009, walter_rapid_2017, heinsoo_rapid_2018} and semiconductor spin \cite{zheng_rapid_2019, schaal_fast_2020} qubit readout in quantum information processing. In circuit quantum electrodynamics (cQED), near-quantum-limited amplifiers enable fast high-fidelity readout and have helped achieve numerous scientific advances, such as the observation \cite{vijay_observation_2011} and reversal \cite{minev_catch_2019} of quantum jumps, the ``break-even" point in quantum error correction \cite{ofek_extending_2016, hu_quantum_2019,campagne-ibarcq_quantum_2020}, and quantum supremacy or quantum advantage \cite{google_quantum_2019}. Josephson traveling-wave parametric amplifiers (JTWPAs) \cite{obrien_resonant_2014, macklin_a_2015, white_traveling_2015} with several gigahertz of bandwidth and a dynamic range of $P_{1\mathrm{dB}}$ approxmately $-100\,$dBm are widely used as preamplifiers in microwave quantum experiments. While near-ideal intrinsic quantum efficiency (QE) has been achieved with Josephson parametric amplifiers (JPAs) \cite{yurke_observation_1989, castellanos-beltran_widely_2007, yamamoto_flux_2008, bergeal_phase_2010, roy_broadband_2015}, the behavior of which is well understood \cite{eichler_controlling_2014,boutin_effect_2017}, the best reported intrinsic QE of JTWPAs remains at least $20\%$ below that of an ideal phase-preserving amplifier, despite several independent implementations  \cite{planat_photonic_2020, ranadive_a_2021}.
Although such reductions in intrinsic QE are commonly attributed almost entirely to dielectric losses, detailed noise characterization suggests that an unknown noise mechanism is dominant \cite{macklin_a_2015}. Such a yet to be identified noise source will limit the best achievable readout fidelity and speed, and eventually hinders the realization of broadband ideal quantum measurements which are critical to continuous quantum error correction \cite{ahn_continuous_2002, ahn_quantum_2003},
quantum feedback control \cite{wang_feedback_2001,tornberg_high_2010, vijay_stabilizing_2012,naghiloo_information_2018}, and ultrasensitive parameter estimation \cite{gammelmark_bayesian_2013}. Consequently, an outstanding, yet unanswered, question is: can a wideband parametric amplifier achieve near-ideal quantum efficiency?

In this work, we quantitatively identify the sidebands as the dominant noise mechanism in existing TWPAs using the multimode, quantum input-output theory framework presented here, which also models propagation loss quantum mechanically. We then offer an affirmative answer to the previous question by introducing Floquet mode amplifiers, a new class of broadband amplifiers that encode information in Floquet modes and effectively eliminate coherent information leakage. A Floquet mode JTWPA can achieve $>\!20\,$dB gain and a QE of $\eta/\eta_{\mathrm{ideal}} > 99.9\%$ of the quantum limit over an instantaneous $3\,$dB bandwidth of approximately $6.5\,$GHz. Importantly, Floquet mode TWPAs strongly suppress the nonlinear forward-backward-mode couplings, which dominate the signal reflection in conventional homogeneous critical current TWPAs that are well impedance matched. Floquet mode TWPAs are thus genuinely directional and can minimize signal reflection to $<\!-25\,$dB over the full amplifying bandwidth, making them integrable with qubits without commercial isolators and potentially enabling near-perfect full-chain measurement efficiency.
Additionally, Floquet mode TWPAs also offer the practical advantages of convenient interfacing using bare frequency modes and insensitivity to out-of-band impedance environment, which strongly suppresses gain ripples. We predict a QE of $\eta/\eta_{\mathrm{ideal}}>99.9\%$ to be realistically achievable using a fabrication process with a dielectric loss tangent $\tan\delta\sim\!10^{-6}$, typical of qubit fabrication.

Although here we illustrate the new Floquet mode amplifier using a microwave JTWPA design, it is worth noting that it can be readily applied to any traveling-wave-style parametric amplifiers such as the kinetic inductance traveling-wave amplifiers (KITs) \cite{eom_a_2012, vissers_low_2016, malnou_three_2021}. We anticipate that Floquet mode TWPAs will help advance various information-critical applications in metrology and quantum information processing,enabling the longstanding goal of broadband  high-sensitivity dark-matter searches \cite{bartram_dark_2021} and paving the way for scalable fault-tolerant quantum computing by enabling fast, multiplexed qubit readout below the surface code error threshold \cite{fowler_high_2009}.

\section{Multimode Dynamics} \label{sec:multimode}

The uncertainty principle of quantum mechanics requires any linear phase-preserving amplifier to add at least approximately a half quantum of noise referred to the input at high gain \cite{haus_quantum_1962, caves_quantum_1982,clerk_introduction_2010}. For an ideal two-mode parametric amplifier, the idler mode, described by the creation operator $\hat{a}^\dagger(\omega_i)$, acts as the coherent ``reservoir" that injects the minimal added noise to the signal $\hat{a}(\omega_s)$ to preserve the bosonic commutator relations at the output. At a power gain $G\geq1$ in linear units, the ideal quantum efficiency is commonly defined as \cite{boutin_effect_2017} 

\begin{figure}[bhp]
\centering
\includegraphics[width=0.98 \linewidth]{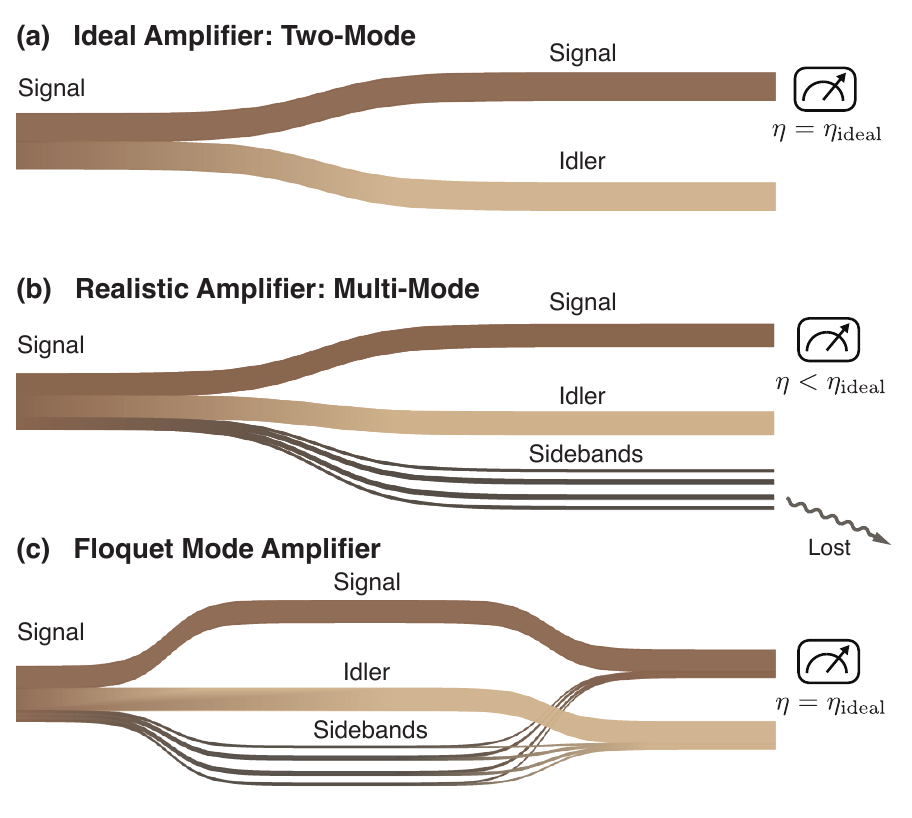}
\caption{{A schematic illustration of the information flow in different types of parametric amplifiers.} (a) Ideal two-mode amplifiers preserve information perfectly, albeit halving the measurement strength as required by the uncertainty principle. (b) Realistic TWPAs are multimode and a portion of the information leaks into the unmeasured sidebands. (c) Floquet mode TWPAs, proposed in this work, can coherently recover the leaked information back into the signal and idler and reach quantum-limited performance, despite their multimode behavior.} 
\label{fig:ampcartoon}
\end{figure}

{
\begin{equation}
    \eta_{\mathrm{ideal}}(G) = \frac{\abs*{\Delta \hat{a}_{\mathrm{in}}(\omega_s)}^2}{\abs*{\Delta \hat{a}_{\mathrm{out}}(\omega_s)}^2/G} = \frac{1}{2-1/G}, \label{eqn:idealqe}
\end{equation}
}

\noindent where $\hat{a}_{\mathrm{in (out)}}(\omega_s)$ is the input (output)  annihilation operator of the signal, and $\abs*{\Delta \hat{a}}^2 \equiv \ev**{\{\hat{a},\hat{a}^\dagger\}}/~2 - \langle \hat{a}^\dagger \rangle \langle \hat{a} \rangle$ is the mean-square fluctuation of operator $\hat{a}$ \cite{caves_quantum_1982}. It is worth emphasizing that $\eta_{\mathrm{ideal}}$ corresponds to the standard quantum limit even though it is only approximately $50\%$ at high gain $G$. No information is lost in the process \cite{clerk_introduction_2010} because the idler is not correlated with any other unmeasured degrees of freedom. Such ideal two-mode amplifiers, albeit halving the measurement strength, can preserve information perfectly as illustrated in \cref{fig:ampcartoon}\textcolor{blue}{(a)}.

\begin{figure}[b]
\centering
\includegraphics[width=0.98 \linewidth]{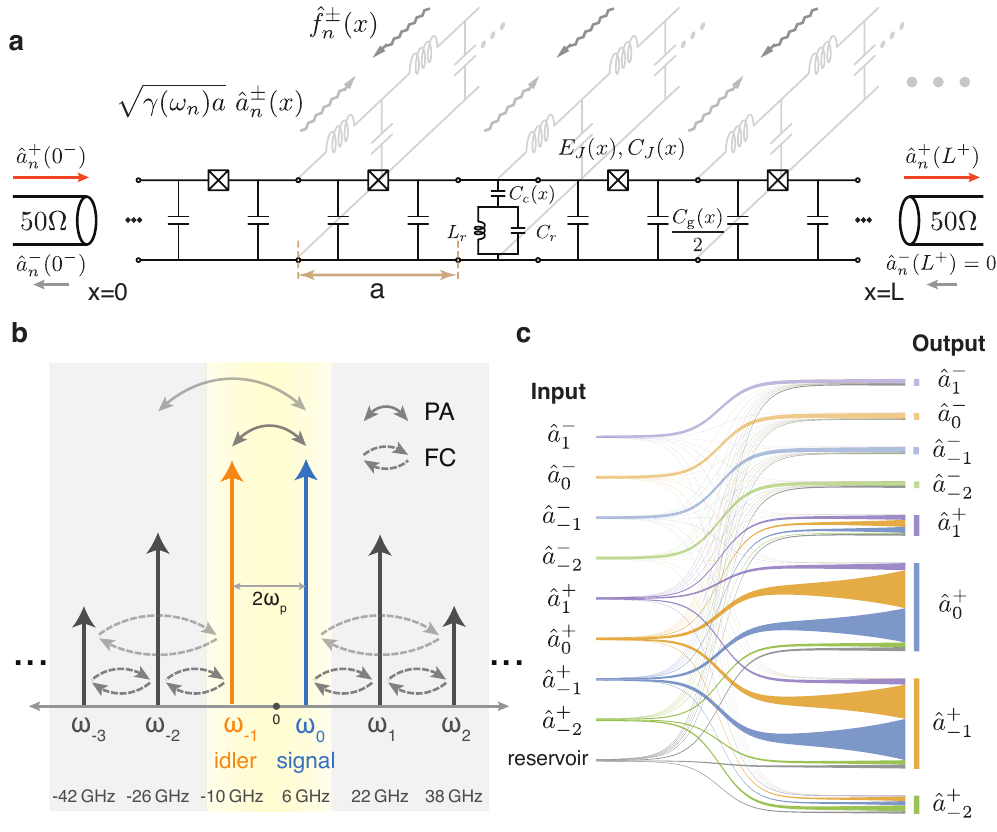}
\caption{The TWPA model and the multimode dynamics. (a) The circuit diagram of a generic resonantly phase-matched JTWPA terminated by $50\,\Omega$ linear transmission line ports at $x=0$ and $x=L$, where $L$ is the device length in unit cells (of length $a$). The unit cell at $x$ consists of a pair of ground capacitors $C_g(x)/2$ and a Josephson junction with junction energy $E_J(x) = \phi_0 I_0(x)$ and capacitance $C_J(x)$, in which $\phi_0 = \Phi_0 / (2\pi)$ is the reduced flux quantum, and $I_0(x)$ is the junction critical current. Phase-matching resonators with inductance $L_r$, capacitance $C_r$, and coupling capacitance $C_c(x)$ are inserted periodically for phase matching. The dielectric loss is modeled by a series of lossless transmission line ports in light gray, the scattering parameters of which are determined by the loss rate $\gamma(\omega)$. $f_n^{\pm}(x)$ are the uncorrelated noise operators introduced to the system by distributed loss. (b) A two-sided frequency spectrum representation of a typical degenerate four-wave mixing TWPA. The signal and idler frequencies are denoted as $\omega_0$ and $\omega_{-1}$, respectively. The solid and dashed arrows indicate the parametric amplification and frequency conversion processes, respectively. The example frequency scale corresponds to a $6$-GHz signal and a $8$-GHz pump. (c) A Sankey-like illustration of the output operator or noise decomposition of a multimode TWPA constructed from the scattering matrices $\mathbb{S}_0$ and $\mathbb{S}_n(x)$. The relative width of each ``water flow" from source node $i$ to target node $j$ represents the operator or noise composition of the input mode $\hat{a}_i$ in the output mode $\hat{a}_j$.} 
\label{fig:setup}
\end{figure}

Practical TWPAs, however, are intrinsically multimode because their large bandwidth allows for a spectrum of sidebands to propagate simultaneously. Sidebands have been a longstanding problem for all types of wideband TWPAs \cite{Sakuraba_extension_1963, mcKinstrie_quantum_2005}. The same pump tone providing the large signal gain necessary to suppress the excess noise of downstream amplifiers also unavoidably induces the sideband couplings, leading to non-negligible noise and information leakage as illustrated in \cref{fig:ampcartoon}\textcolor{blue}{(b)}. Although such sideband couplings are also present in our proposed Floquet mode amplifiers, we show later that Floquet mode amplifiers are nevertheless still able to circumvent this issue and reach the quantum limit over their broad working bandwidth. This can be intuitively understood either as the Floquet mode amplifiers coherently recovering the leaked information back into the signal and idler mode in the bare frequency picture [\cref{fig:ampcartoon}\textcolor{blue}{(c)}] or as encoding information instead in the collective Floquet modes of the driven system and adiabatically mode matching to the bare frequency modes at input and output for convenient interfacing in the Floquet-mode picture.

The various two-pump-photon parametric interactions responsible for the coherent information leakage in a typical degenerate four-wave mixing (4WM) TWPA can be conveniently visualized using the two-sided frequency spectrum in \cref{fig:setup}\textcolor{blue}{(b)}. In this representation, the idler mode has a defined negative frequency of $\omega_{-1} = \omega_s - 2\omega_p = -\omega_i$ in recognition of the relation $\hat{a}(-\omega_i)=\hat{a}^\dagger(\omega_i)$ \cite{roy_introduction_2016}. Similarly, the frequencies of the gray-colored sideband modes $\hat{a}_n = \hat{a}(\omega_n)$ are specified by $\omega_n = \omega_s + 2n\omega_p$, in which the integer mode index $n$ can be either negative or positive. Following this convention, 4WM processes couple only adjacent modes that are spaced by $2\omega_p$, whereas six-wave-mixing (6WM) processes can instead couple modes separated by up to two spacings and so forth. Frequency conversion (FC) and parametric amplification (PA) processes can be distinguished by whether the frequencies of their two interacting modes possess the same or opposite signs, respectively. Analogously to a qubit, information can coherently leak out from the ``computational subspace" of the signal and the idler in the yellow-shaded region of \cref{fig:setup}\textcolor{blue}{(b)} into the sidebands, in addition to incoherent losses from radiation or dissipation. The effects of sidebands on phase matching and hence gain dynamics have been studied in KITs and JTWPAs \cite{chaudhuri_simulation_2015, erickson_theory_2017, dixon_capturing_2020} but the connection between sidebands and the noise performance has not been recognized, potentially due to the lack of a systematic rigorous quantum framework.

To quantify the effects of both sideband leakage and propagation loss on the QE of TWPAs, we develop a multimode quantum input-output theory framework, as illustrated in \cref{fig:setup}\textcolor{blue}{(a)}. The circuit design exemplified in the figure is similar to, but more general than, that of a typical resonantly phase-matched JTWPA \cite{obrien_resonant_2014, macklin_a_2015}, as it allows the device to be inhomogeneous and the circuit parameters to have a spatial dependence. The propagation loss is modeled quantum mechanically with a series of distributed, lossless, and semi-infinite transmission line ports. Dissipation and their associated fluctuations can then be cast as the coherent scattering into and from these transmission lines, the frequency-dependent scattering parameters of which are set by the loss rate $\gamma(\omega)$ \cite{caves_quantum_1987, jeffers_quantum_1993, houde_loss_2019}. Furthermore, we extend the beam-splitter model such that our model can now work for any generic second-order equation of motion and properly account for the loss and interactions among the forward and backward modes without taking the slowly varying envelope approximation (for details, see \cref{app:spatialeom}). In addition, our model can correctly account for impedance mismatch at the boundaries, insertion loss, and nonlinear processes from arbitrary orders of pump nonlinearities (4WM, 6WM, $\cdots$). 

Under the stiff-pump approximation, the multimode system can be linearized around a strong, classic pump with a dimensionless amplitude $I_{pn}(x) = I_p(x) / I_0(x)$, where $I_{p}(x)$ and $I_0(x)$ are the pump current and the junction critical current at $x$, respectively. In the continuum limit ($\abs{k_{s,i}}a \ll 1$), the quantum spatial dynamic equations in the frequency domain can be derived from the Heisenberg equations and written in block matrix form as follows:

{
\begin{align}
    \dv{x} \begin{pmatrix} 
 \vec{\mathrm{A}}^+ (x)\\  \vec{\mathrm{A}}^-(x)
\end{pmatrix} &=  \mathbb{K}(x) \begin{pmatrix}  \vec{\mathrm{A}}^+ (x)\\  \vec{\mathrm{A}}^-(x)
\end{pmatrix} \nonumber \\
 &\quad + \frac{1}{2}\begin{pmatrix} -{\Gamma}(x) & \mathbf{0}_m \\ \mathbf{0}_m & {\Gamma}(x)\end{pmatrix}
\begin{pmatrix} 
 \vec{\mathrm{A}}^+(x) \\  \vec{\mathrm{A}}^-(x)
\end{pmatrix} \nonumber \\  
&\quad  +\begin{pmatrix} \sqrt{{\Gamma}(x)} & \mathbf{0}_m  \\ \mathbf{0}_m  & \sqrt{{\Gamma}(x)} \end{pmatrix} \begin{pmatrix} 
 \vec{\mathrm{F}}^+(x) \\  \vec{\mathrm{F}}^-(x)
\end{pmatrix}, \label{eqn:eom}
\end{align}
}

\noindent in which $\vec{\mathrm{A}}^{\pm}(x) = [\cdots, ~\hat{a}^{\pm}_n(x),  ~\cdots]^{\mathrm{T}}$ are the forward- and backward-propagating operator vectors, $\vec{\mathrm{F}}^{\pm}(x)$ are, similarly, the forward and backward noise operator vectors, $\Gamma(x) = \mathrm{diag}\left[\cdots, \gamma(\omega_n),  \cdots \right]$ is the diagonal loss-rate matrix, and $\mathbf{0}_m$ is the $m\times m$ zero matrix, with $m$ being the number of frequency modes considered. The field ladder operators $\hat{a}^{\pm}_n(x)$ satisfy the commutation relations \cite{roy_introduction_2016}

{
\begin{align}
    \comm{\hat{a}^{d_1}_{n_1}(x)}{\hat{a}^{d_2\dagger}_{n_2}(x')} &= \mathrm{sgn}(\omega_{n_1})\delta_{n_1,n_2}\delta_{d_1,d_2}\delta(x-x'), \label{eqn:commrelation}
\end{align}
}

\noindent in which the superscripts $d_1, d_2\in\{+,-\}$ denote the forward-(+) or backward-(-) propagating modes and $\mathrm{sgn}(\omega)$ is the sign function. The noise operators $\hat{f}_n^{\pm}(x)$ follow a commutation relation that is similar to  \cref{eqn:commrelation}.

The multimode coupling matrix $\mathbb{K}(x)$ in the first term of \cref{eqn:eom} is positive definite and captures the various PA, FC, and forward-backward coupling processes (see \cref{app:spatialeom}), whereas the second and third terms describe the dissipation and the associated fluctuation from material loss respectively. 

By solving \cref{eqn:eom} and applying the proper boundary conditions at $x=0$ and $x=L$, we can relate the input and output bosonic modes $\vec{\mathrm{A}}_{\mathrm{in}}$ and  $\vec{\mathrm{A}}_{\mathrm{out}}$ in the transmission lines ports terminating the TWPA by (see \cref{app:inputputtheory})

{
\begin{equation} 
    \vec{\mathrm{A}}_{\mathrm{out}} = \mathbb{S}_0 \vec{\mathrm{A}}_{\mathrm{in}} + \int_0^L dx \left( \mathbb{S}_n(x) \mathrm{\Gamma}^{1/2}(x) \vec{\mathrm{F}}(x) \right),\label{eqn:scatteringrelation}
\end{equation}
}

\begin{figure}[htp!]
\noindent
\includegraphics[width=1\linewidth]{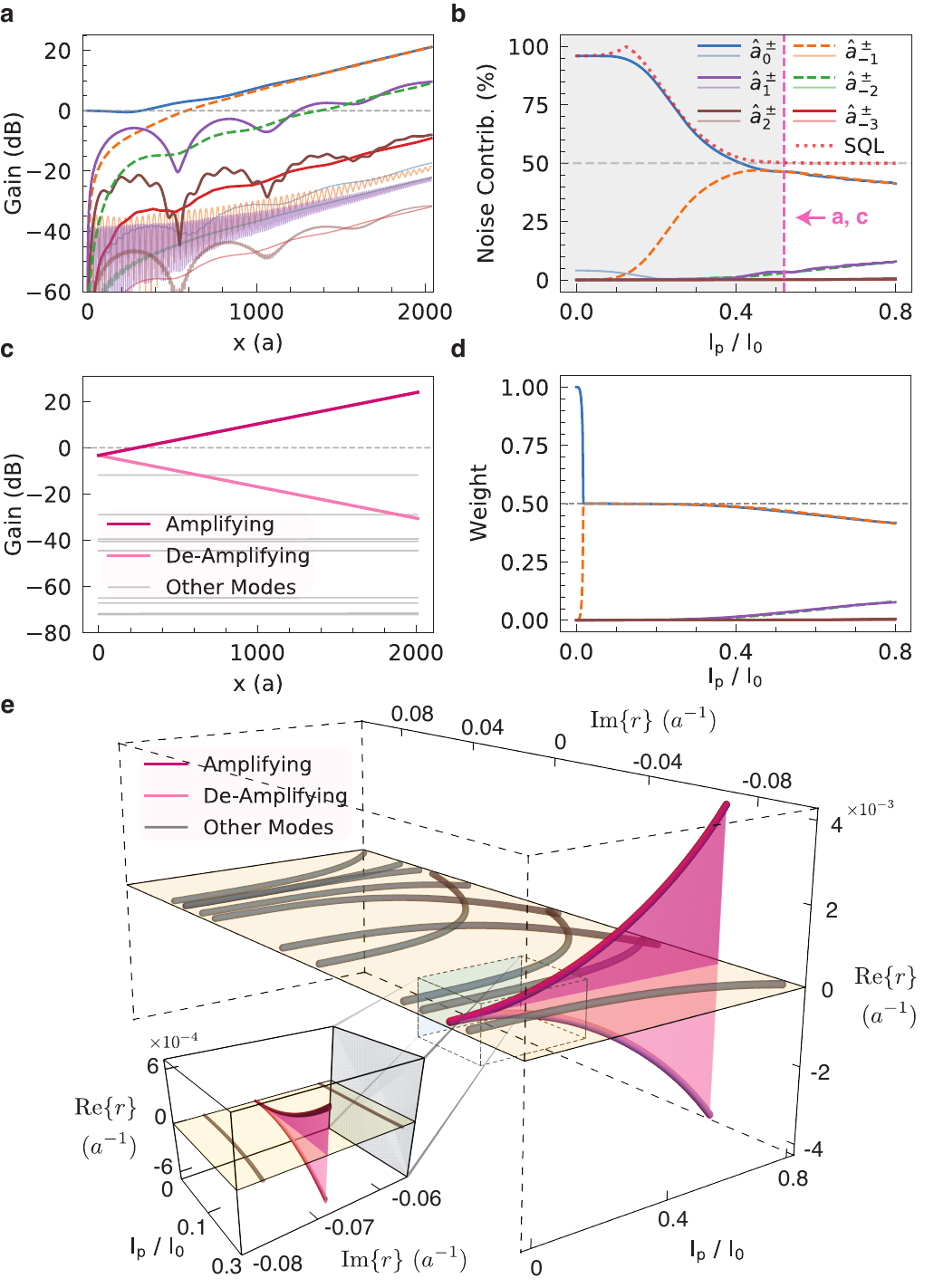}
\caption{A comparison between the frequency- and Floquet-basis representations: panels (a), (b), and (d) share the same legend. Forward ($\hat{a}_n^+$) and backward ($\hat{a}_n^-$) frequency modes are plotted in the same color but with different line styles ($\hat{a}_n^+$ in thick opaque lines and $\hat{a}_n^-$ in thin semitransparent lines, respectively). The modes $\hat{a}_{-1}^+$ and $\hat{a}_{-1}^+$ overlap significantly with $\hat{a}_{0}^+$,$\hat{a}_{1}^+$ and are thus plotted in dashed lines for visual clarity. (a) The internal spatial dynamics in the frequency basis when a forward signal is injected and driven by $I_{pn} = I_p / I_0 = 0.52$. (b) The output signal noise decomposition as a function of the normalized pump current $I_{pn}$ with $Z_0=50\,\Omega$ port impedance. The region with $<\!20\,$dB gain is shaded in light gray.
Panels (a) and (c) correspond to the vertical pink dashed line denoted by an arrow. (c) The same spatial dynamics as in (a) but plotted in the Floquet mode basis. All but the amplifying and deamplifying Floquet modes are stable and colored in light gray. (d) The frequency-mode decomposition of the amplifying Floquet mode $\hat{Q}_a$ as a function of $I_{pn}$.The bifurcation point of the signal and idler components at $I_{pn}\approx0.02$ denotes the boundary of the regions of stability and instability. (e) The Floquet characteristic exponents $r_{\alpha}$ plotted as a function of $I_{pn}$ (x axis). The y-z plane corresponds to the complex plane of $r_\alpha$. The yellow-shaded x-z plane corresponds to zero gain coefficient $g_\alpha = \Re{r_\alpha}=0$ and encloses all other sidebandlike Floquet modes. The inset shows that $\hat{Q}_a$ and $\hat{Q}_d$ do not intersect with other Floquet modes. All panels are generated using $m=6$ frequency modes and we choose the pump wave vector expression $k_p(I_{pn}) = 0.08195 + 1.1601\times10^{-2} I_{pn}^2 - 4.996\times10^{-3}I_{pn}^3$ to maximize the signal gain above bifurcation.
\label{fig:floquetbasis}}
\end{figure}

\noindent where $L$ is the device length in unit cells and $\mathbb{S}_0$ and $\mathbb{S}_n(x)$ are the $2m \times 2m$ multimode and noise scattering matrices that capture the effects of sideband couplings and dissipation on the output signal, respectively. Together, they are sufficient to calculate the full quantum statistics of the output modes. The quantum efficiency $\eta$ of a general multimode TWPA with dissipation can thus be calculated from $\mathbb{S}_0$ and $\mathbb{S}_n(x)$ as 

\begin{widetext}
  {
\begin{align}
    &\eta(G) = \eta(\abs{\mathbb{S}_{{0},ss}}^2) = \frac{\abs*{\Delta \hat{a}_{\mathrm{in}}(\omega_s)}^2}{\abs*{\Delta \hat{a}_{\mathrm{out}}(\omega_s)}^2/\abs{\mathbb{S}_{{0},ss}}^2} \nonumber \\
    &=\frac{\abs{\mathbb{S}_{{0},ss}}^2 \abs*{\Delta \vec{\mathrm{A}}_{\mathrm{in},s}}^2 }{\sum_k{\abs{\mathbb{S}_{{0},sk}}^2 \abs*{\Delta \vec{\mathrm{A}}_{\mathrm{in},k}}^2+ \int_0^L{\gamma(\omega_k)\abs*{\mathbb{S}_{{n},sk}}^2 \abs*{\Delta \vec{\mathrm{F}}_{k}(x)}^2 }dx} } \label{eqn:qeexpr},
\end{align}
}
\end{widetext}

\noindent where $\vec{\mathrm{A}}_k$ denotes the k-th operator of the vector $\vec{\mathrm{A}} \in \{\vec{\mathrm{A}}_{\mathrm{in}}, \vec{\mathrm{F}}(x)\}$,  $s$ is the index of the forward signal mode $\hat{a}_0^+$, and $\abs*{\mathbb{S}_{{n},sk}}^2$ is the signal power gain $G$. Note that \cref{eqn:qeexpr} can be mapped to the usual form $\eta = (1/2) / [(1/2)+\mathcal{A}]$ with Cave's added noise number $\mathcal{A} = (\sum_{k}\abs{\mathbb{S}_{{0},sk}}^2 + \int_0^L{\gamma(\omega_k)\abs*{\mathbb{S}_{{n},sk}}^2}dx )/\abs{\mathbb{S}_{{0},ss}}^2 - 1$, in the case when all input and environmental states have the minimum vacuum noise of  $\abs*{\Delta\vec{\mathrm{A}}_{\mathrm{in,k}}}^2 = \abs*{\Delta\vec{\mathrm{F}}_{\mathrm{k} }(x)}^2 = 1/2$.

\Cref{fig:setup}\textcolor{blue}{(c)} pictorializes the dynamics of a multimode TWPA from the perspective of added noise using information from the scattering matrices $\mathbb{S}_0$ and $\mathbb{S}_n(x)$. A ``water flow" from source mode $i$ to target mode $j$ can be equally interpreted as the operator or noise composition of input mode $i$ in the output mode $j$. The noise of component $i$ in output $j$ is proportional to the width of the $i\rightarrow j$ path. The quantum efficiency $\eta$ therefore corresponds to the weight (width) of the input signal noise in the total output signal noise (width). In the case of an ideal two-mode parametric amplifier, the output signal would have all of its in-flow (noise) coming from only the input signal and idler, signifying quantum-limited noise performance. In contrast, as is shown for a practical TWPA, an additional, non-negligible portion of the output signal noise instead comes from the input sideband modes and the reservoir of fluctuation operators, degrading the QE.

\section{The Floquet-Basis Picture} \label{sec:floquetpicture}

We now numerically simulate and visualize the spatial dynamics inside a JTWPA, the circuit parameters of which are similar to those in Ref. \cite{macklin_a_2015}, except that here we assume $\tan\delta=0$ here (see \cref{tab:floquetbasisparams} in \cref{app:circuitparams}). In \Cref{fig:floquetbasis}\textcolor{blue}{(a)}, a unit of forward signal $\hat{a}_0^+$ is injected from $x=0$ with $I_{pn} \equiv I_p / I_0 = 0.52$, which is chosen to produce approximately $20\,$dB gain in $L=2037$ cells. Although still being amplified, the forward signal and idler are continuously being converted to the sidebands $\hat{a}_{1}^+$ and $\hat{a}_{-2}^+$, leading to the sideband amplification and information leakage.

\Cref{fig:floquetbasis}\textcolor{blue}{(b)} plots the noise decomposition of the amplified output signal $\hat{a}_0^+(x=L^+)$ in the $50\,\Omega$ output port as a function of $I_{pn}$. The device length is fixed at $L=2037$, such that the signal gain increases nearly monotonically with $I_{pn}$ before the onset of parametric oscillations. From \cref{eqn:qeexpr}, the QE $\eta$ can be usefully interpreted as the ratio of the amount of noise from the original signal $\hat{a}_0^+$ to the total output signal noise. QE therefore maps exactly to the signal contribution (in solid blue) in \cref{fig:floquetbasis}\textcolor{blue}{(b)}, which decreases with the $I_{pn}$ due to an elevated sideband contribution. This quantitatively accounts for the unknown QE reduction in Ref. \cite{macklin_a_2015} and provides numerical evidence that the sidebands are indeed a significant noise source in JTWPAs.

Floquet theory \cite{teschl_ordinary_2012} provides invaluable insights into the noise performance of TWPAs. Floquet modes are a set of solutions for a periodically driven system that forms a complete orthonormal basis. Each Floquet mode can be expressed as $\hat{Q}_\alpha(x) = e^{r_\alpha x} \hat{\alpha}(x)$, where $\hat{\alpha}(x)$ is spatially periodic and $r_\alpha$ is the complex Floquet characteristic exponent of Floquet mode $\alpha$. For a homogeneous lossless TWPA described by \cref{eqn:eom}, the solution can be expressed in the form of $\vec{\mathrm{A}}(x) = \mqty[ \vec{\mathrm{A}}^+(x) & \vec{\mathrm{A}}^-(x)]^T = \Pi(x) \vec{A} (0)$, in which the transfer matrix $\Pi(x)$ can be written in the form of $\Pi(x) = \mathbb{P}(x)\mathrm{exp}(x\mathbb{Q})$ according to the Floquet theorem and $\mathbb{P}(x)$ has the same periodicity as $\mathbb{K}(x)$. We can thus transform the frequency-basis vector $\vec{\mathrm{A}}(x)$ into the Floquet-mode basis $\vec{\mathrm{Q}}(x)$ via (for details, see \cref{app:floquet})

{
\begin{equation}
    \vec{\mathrm{Q}}(x) = \left( \mathbb{P}(x)\mathbb{V} \right)^{-1}\vec{\mathrm{A}}(x) \label{eqn:fqbasis}
\end{equation}
} 
\noindent where $\mathbb{V}$ is the orthonormal basis of matrix $\mathbb{Q}$ such that $\mathbb{Q}=\mathbb{V}\Lambda\mathbb{V}^{-1}$ with $\Lambda=\mathrm{diag}(\cdots, r_\alpha, \cdots)$ containing the Floquet exponents. \Cref{fig:floquetbasis}\textcolor{blue}{(c)} shows the exact same spatial evolution as in \cref{fig:floquetbasis}\textcolor{blue}{(a)} but in the Floquet-mode basis. A unit of forward input signal $\hat{a}_0^+(0)$ is injected and projected into a collection of Floquet modes, each of which then propagates separately with a distinct complex propagation constant $r_{\alpha}$. Notably, there is only one amplifying and one deamplifying Floquet mode $\hat{Q}_a$ and $\hat{Q}_d$, which can be understood as the antisqueezing and squeezing quadratures of the signal-idler-like mode as shown in \cref{fig:floquetbasis}\textcolor{blue}{(d)}. All the other sidebandlike Floquet modes, colored in gray, remain constant in space (for details, see \cref{app:floquet}). This suggests an alternative view to the sideband-induced excess noise and coherent information leakage: they result from the mode mismatch between the bare frequency modes of the input and output ports and the collective Floquet modes of a driven TWPA.

At $G\gg 1$, the amplifying Floquet mode $\hat{Q}_a$ dominates the gain and noise performance of a TWPA. \Cref{fig:floquetbasis}\textcolor{blue}{(d)} plots the frequency-mode decomposition of $\hat{Q}_a$ as a function of $I_{pn}$. The Hopf bifurcation point of the signal and idler components at $I_{pn}\approx0.02$ marks the transition of the system from the region of stability to instability (amplification), the exact position of which is dependent upon the phase mismatch. The mixing of signal and idlers and the bifurcation of complex-conjugate pairs $r_a$ and $r_d$ [\cref{fig:floquetbasis}\textcolor{blue}{(e)}] implies the standard quantum limit for phase-preserving amplifiers: signal-idler mixtures of different relative phases are split into the amplifying and deamplifying Floquet modes and thus cannot be all noiselessly amplified at the same time. The gain coefficient $g_a=\Re{r_a}$ of $\hat{Q}_a$ [\cref{fig:floquetbasis}\textcolor{blue}{(e)}] increases monotonically with $I_{pn}$ and $\hat{Q}_a$ mixes in more sideband modes; in particular, $\hat{a}_1$ and $\hat{a}_{-2}$. The remarkable resemblance between \cref{fig:floquetbasis}\textcolor{blue}{(b)} and \cref{fig:floquetbasis}\textcolor{blue}{(d)} in the region of large signal gain shows the usefulness of the Floquet basis in understanding the TWPA noise performance. An increased sideband weight in $\hat{Q}_a$ means that a larger portion of the sideband vacuum fluctuations incident upon a TWPA would be projected into the amplifying Floquet mode and would then subsequently generate more noise power at the signal and idler frequencies. The dependence of the $\hat{Q}_a$ mode mixture on $I_{pn}$ also sheds light on the experimental observation that the peak signal-to-noise ratio (SNR) improvement often does not coincide with the largest signal gain as a function of pump power \cite{macklin_a_2015}.

\begin{figure*}[ht]
\noindent
\includegraphics[width=0.98\linewidth]{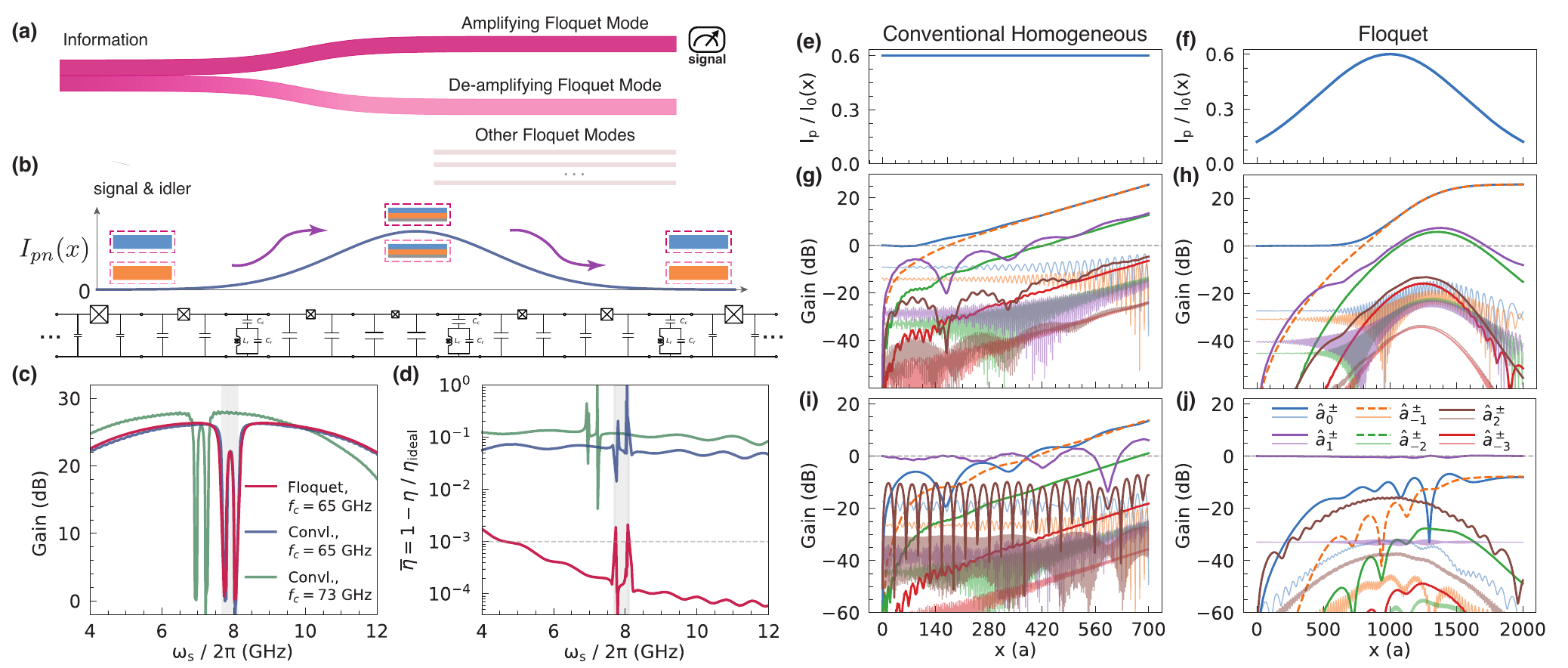}
\caption{The adiabatic Floquet mode scheme and the Floquet mode JTWPA performance. (a) A schematic illustration of the information flow in a Floquet mode TWPA. It is equivalent to \cref{fig:ampcartoon}\textcolor{blue}{(c)} but in the Floquet basis. (b) The Floquet mode TWPA circuit design and the effective drive-amplitude profile $I_{pn}(x)$. The ground capacitance $C_g(x)$ is varied accordingly with the junction energy $E_J(x)$ to keep the nominal characteristic impedance constant. The effective pump currents at both ends of the device are ideally zero. (c) The gain spectrum of the 2000-cell Floquet (red) and the 700-cell conventional homogeneous (blue) scheme with similar dispersion with $\omega_p/(2\pi)=7.875\,$GHz and $I_{pn} = 0.6$. The simulated gain of the device in Ref. \cite{macklin_a_2015} with 2037 cells is also plotted in green for comparison. (d) The amplifier quantum inefficiency $\bar{\eta} =1-\eta/\eta_{\mathrm{ideal}}$ spectrum of the conventional and Floquet mode scheme. The simulated $\bar{\eta}$ of the device in Ref. \cite{macklin_a_2015} is also plotted for comparison. (e), (f) The effective normalized pump-current profile $I_{pn}(x)$ as a function of the position of the homogeneous and the proposed Floquet TWPA respectively. (g), (h) Internal field profiles of the homogeneous and Floquet mode TWPA respectively interfaced by the $50\,\Omega$ transmission line boundaries when the forward signal mode $\hat{a}_0^+$ (solid blue) is injected. (i), (j) Internal field profiles of the homogeneous and Floquet mode TWPA, respectively, when the forward sideband vacuum mode $\hat{a}_1^+$ (solid purple) is injected. Throughout this figure, the dielectric loss tangent is assumed to be zero. }
\label{fig:floquetdesign}
\end{figure*}

\Cref{fig:floquetbasis}\textcolor{blue}{(e)} plots in three dimensions the complex Floquet characteristic exponents $r_{\alpha}$ as a function of $I_{pn}$ (x axis). All but the amplifying and deamplifying Floquet modes stay within the plane of $g_\alpha = \Re{r_\alpha} = 0$ throughout and are stable, whereas the gain coefficient magnitudes of $\hat{Q}_a$ and $\hat{Q}_d$ increase with $I_{pn}$ as expected. It is important to point out that $r_a$ and $r_d$, the complex eigenvalues of $\hat{Q}_a$ and $\hat{Q}_d$ respectively, do not ever intersect with those of any other Floquet modes at all values of $I_{pn}$, as made clear in the inset of \cref{fig:floquetbasis}\textcolor{blue}{(e)}. The existence of a gap between $r_a$ $(r_d)$ and the rest of the spectrum will be crucial to the Floquet mode amplifiers introduced in \cref{sec:floquettwparesults}. 

\section{The Floquet Mode Amplifiers} \label{sec:floquettwparesults}

We now introduce the Floquet mode amplifiers, which can effectively eliminate the aforementioned sideband-induced noise and approach the standard quantum limit simultaneously over the broad operating bandwidth. As alluded to earlier in \cref{sec:multimode} and depicted in \cref{fig:floquetdesign}\textcolor{blue}{(a)},  the principal idea behind Floquet mode amplifiers is that they are effective two-``Floquet mode" amplifiers, as information is exclusively encoded in the two instantaneous collective amplifying and deamplifying Floquet modes, $\hat{Q}_a(x)$ and $\hat{Q}_d(x)$, of the driven system. Furthermore, we adiabatically transform the instantaneous Floquet modes inside the amplifier with a spatially varying dimensionless drive-amplitude profile $I_{pn}$ as illustrated in \cref{fig:floquetdesign}\textcolor{blue}{(b)}. This allows us to mode match the information-carrying Floquet modes to the input and output bare single frequency modes $\hat{a}^+_0$ for convenient interfacing, as in the case of existing TWPAs. $I_{pn}(x)$ are ideally set to near zero at the boundaries to perfectly mode match the signal and idler modes $\hat{a}_0^+$ and  $\hat{a}_{-1}^+$ in the linear input and outputs to the instantaneous $\hat{Q}_a(x)$ and $\hat{Q}_d(x)$. Near the center of the amplifier, $I_{pn}(x)$ is adiabatically ramped up to significantly amplify the signal within a reasonable device length. Note that the increased sideband mixture in $\hat{Q}_a(x)$ and $\hat{Q}_d(x)$ in the middle does not contribute to additional noise in this scheme, because they will eventually be adiabatically transformed back to  $\hat{a}_0^+$ and  $\hat{a}_{-1}^+$ as $I_{pn}(x)$ ramps down to near zero again in the end. As a result, from the view outside of the device, the signal and idler modes are effectively decoupled from the various sidebands and Floquet mode amplifiers can therefore approach quantum-limited noise performance.

In practice, we can tailor the desired spatial profile of $I_{pn}(x)$ from a constant input pump current $I_p$ by instead varying the junction critical current $I_0(x)$ and ground capacitance $C_g(x)$ as shown in \cref{fig:floquetdesign}\textcolor{blue}{(b)}. Here, we consider the case in which the spatially varying $I_0(x)$ is achieved with varying junction areas but the same plasma frequency using a single fabrication step. We also vary the coupling capacitance $C_c(x)$ such that the coupling strength $C_c(x)/C_g(x)$ of phase-matching resonators (PMRs) remains constant and the phase-matching condition is similar throughout the device (see \cref{app:lagrangianhamiltonian}). In Figs. \ref{fig:floquetdesign}\textcolor{blue}{(e)}-\ref{fig:floquetdesign}\textcolor{blue}{(j)}, we compare the spatial dynamics of a conventional homogeneous critical current design [Figs. \ref{fig:floquetdesign}\textcolor{blue}{(e)}, \ref{fig:floquetdesign}\textcolor{blue}{(g)}, and \ref{fig:floquetdesign}\textcolor{blue}{(i)}] with our proposed Floquet scheme [Figs. \ref{fig:floquetdesign}\textcolor{blue}{(f)}, \ref{fig:floquetdesign}\textcolor{blue}{(h)}, and \ref{fig:floquetdesign}\textcolor{blue}{(j)}]. Both designs use a slightly reduced cutoff frequency $f_c = \omega_c/(2\pi) \sim 65\,$GHz with two junctions in series in one unit cell to reduce the physical device length ($700$ and $2000$ unit cells, respectively) necessary to achieve approximately $25\,$dB gain (for the circuit parameters, see \cref{tab:floquetdesignparams} in \cref{app:circuitparams}). For comparison, the JTWPA in Ref. \cite{macklin_a_2015} has a slightly higher cutoff $f_c = 73\,$GHz and $2037$ cells. Whereas the conventional homogeneous design has a constant drive amplitude of $I_{pn}(x) = I_{pn0} = 0.6$ [\cref{fig:floquetdesign}\textcolor{blue}{(e)}], the instantaneous junction critical current $I_0(x)$  in the adiabatic Floquet scheme constructs a Gaussian profile of $ I_{pn}(x) = I_{pn0}~\exp\left[-(x-\mu)^2/(2\sigma^2)\right]$ [\cref{fig:floquetdesign}\textcolor{blue}{(f)}], which centers at $\mu=L/2=1000$ and has a full width at half maximum (FWHM) of $2\sqrt{2\ln{2}}\sigma = 0.62 L$. This practical choice of $\mathrm{FWHM}$ leads to a small but nonzero $I_{pn}(x)\approx0.1$ at the boundaries, but this still results in nearly ideal quantum efficiency, as it is close to the bifurcation point ($I_{pn}\approx0.05$). Furthermore, here the minimum and maximum junction currents required to achieve an overall dynamic range of approximately $-100\,$dBm are around $3.5\,\mu\mathrm{A}$ and $21.2\,\mu\mathrm{A}$, both of which can be readily fabricated with Lecocq-style junctions \cite{lecocq_junction_2011} or in a niobium-trilayer process \cite{tolpygo_deep_2014},  demonstrating the practicality and robustness of our scheme.

Figures \ref{fig:floquetdesign}\textcolor{blue}{(g)} and \ref{fig:floquetdesign}\textcolor{blue}{(h)} show, respectively, the internal field profiles of the homogeneous and Floquet scheme in the frequency basis when a forward input signal $\hat{a}_0^+$ is injected at $x=0^-$. While the signal and idlers are  amplified by approximately $25\,$dB in both schemes, the Floquet scheme efficiently suppresses the sidebands and the backward modes. Figures \ref{fig:floquetdesign}\textcolor{blue}{(i)} and \ref{fig:floquetdesign}\textcolor{blue}{(j)} show the system response of both schemes when, instead, only the sideband vacuum fluctuation $\hat{a}^+_1(0^+)$ is injected. Whereas the conventional homogeneous design generates a significant amount of added noise at the signal and idler frequencies, the Floquet scheme minimizes the coupling from $\hat{a}^+_{1}$ to $\hat{Q}_a$ and thus suppresses the sideband-induced noise by several orders of magnitude, thereby attaining near-ideal quantum-limited noise performance.

To clearly distinguish the noise performance of different near-quantum-limited amplifiers, we define the amplifier quantum inefficiency as 
{
\begin{equation}
    \bar{\eta}(G) = 1 - \frac{\eta(G)}{\eta_{\mathrm{ideal}}(G)}, \label{eqn:inefficiencydef}
\end{equation}
}
\noindent which signifies the relative difference in the resulting output SNR between a realistic and an ideal phase-preserving amplifier at the same power gain. An ideal preserving amplifier will therefore, by definition, have an inefficiency of $\bar{\eta}=0$, denoting the standard quantum limit. In Figs. \ref{fig:floquetdesign}\textcolor{blue}{(c)} and \ref{fig:floquetdesign}\textcolor{blue}{(d)}, we plot, respectively, the simulated gain and quantum inefficiency spectrum of the proposed Floquet scheme and the conventional homogeneous design. We also include the simulated performance of the experimental device from Ref. \cite{macklin_a_2015} at a similar gain level for comparison. Our multimode quantum model predicts a quantum inefficiency $\bar{\eta}= 0.13$ or $\eta/\eta_{\mathrm{ideal}}=87\%$ for the experimental device assuming no dielectric loss, which is in good agreement with the experimentally extracted value of $\eta/\eta_{\mathrm{ideal}} = 85\%$ in Ref. \cite{macklin_a_2015}. This suggests that our multimode quantum model is able to accurately predict and identify the previously unknown experimentally measured noise mechanism as the sideband-induced noise. For both conventional homogeneous schemes, the quantum inefficiency is still on the order of $10^{-1}$ away from the standard quantum limit due to the additional sideband-induced noise, although a design with a lower cutoff frequency of $65$ GHz (blue curves in Figs. \ref{fig:floquetdesign}\textcolor{blue}{(c)} and \ref{fig:floquetdesign}\textcolor{blue}{(d)}] shows a slight improvement. Notably, sideband effects also manifest themselves in the visible oscillations on the quantum inefficiency or noise spectrum of the conventional homogeneous schemes. Such oscillations in the amplifier added noise have been observed in experiments \cite{malnou_three_2021, ranadive_a_2021}.

In contrast, the Floquet mode TWPA is able to both produce high gain and attain near-ideal QE over a large bandwidth of $6.5\,$GHz (after excluding the band gap due to the phase-matching resonators), well exceeding an octave in $2000$ unit cells. The vanishingly small quantum inefficiency of the adiabatic Floquet design is a direct consequence of the effective decoupling of the signal and idler from the sidebands. The quantum inefficiency $\bar{\eta}$ of the Floquet mode TWPA is shown in \cref{fig:floquetdesign}\textcolor{blue}{(h)} to be smaller than $10^{-3}$ over the full amplifying bandwidth, which is orders of magnitude closer to the quantum limit and can be practically realized.

We note that the broadband signal gain of the Floquet mode TWPA can be further increased if desired, either by driving at a slightly larger $I_{pn}$ \textit{in situ} with a minor decrease in quantum efficiency or with a slightly longer device such that the quantum efficiency remains similar (for details, see \cref{app:gainscaling}). For instance, $\geq 30\,$dB gain can be achieved by driving the discussed Floquet TWPA design at $I_{pn}=0.635$ ($6\%$ increase) or with a longer device of 2100 cells (a $5\%$ increase).

In addition, the upper limit of the dynamic range of TWPA is dominated by pump depletion. The power dependence of the signal gain has the approximate form $G(P_s) = G_0 / (1 + 2G_0P_s/P_p) $ \cite{obrien_resonant_2014}, where $G_0$ is the small signal gain and $P_s$($P_p$) is the signal(pump) power. Because the cutoff frequency of a Floquet mode TWPA is determined by its smallest critical current junctions in the middle, the critical current of all Floquet mode TWPA junctions are no less than those of the corresponding conventional TWPA. Consequently, with the same driving strength $I_{pn}(x)$ and cutoff frequency $f_c$, the pump power $P_p$ and thus the dynamic range of the Floquet TWPA are both larger or equal to those of its conventional TWPA counterpart. The $1$-dB gain compression power of the presented Floquet TWPA design is estimated to be  $P_{1\mathrm{dB}}\sim-100.4\,$dBm, on par with those of conventional TWPAs reported in the literature \cite{esposito_perspective_2021}.

\section{Dielectric Loss, Directionality, and on-Chip Integration} \label{sec:directionality}

\begin{figure}[thp]
\noindent
\includegraphics[width=1\linewidth]{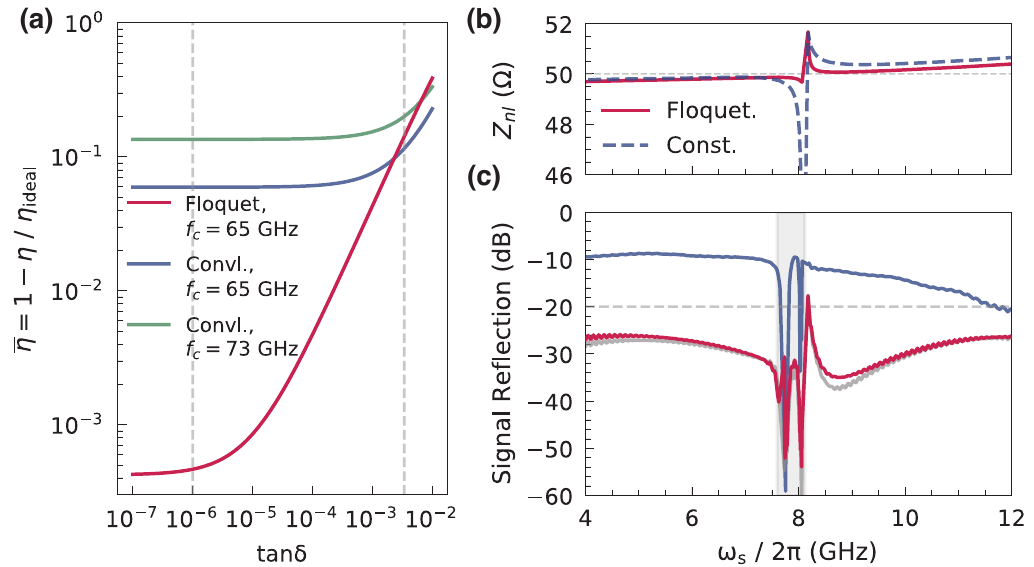}
\caption{The potential of on-chip integrating Floquet mode TWPAs. (a) The quantum inefficiency $\bar{\eta}=1-\eta/\eta_{\mathrm{ideal}}$ of the three amplifier designs in \cref{fig:floquetdesign} as a function of the dielectric loss tangent $\tan\delta$. (b) The nonlinear impedance of the signal as a function of the frequency. (c) The signal reflection as a function of the frequency, where the gray curve corresponds to the signal reflection of the Floquet scheme with nonlinear forward-backward coupling excluded from the equations of motion. $\tan\delta=0$ is assumed in (b) and (c).}
\label{fig:integration}
\end{figure}

We now discuss the nonideality of finite dielectric loss with \cref{fig:integration}\textcolor{blue}{(a)}, in which the quantum inefficiency $\bar{\eta} $ at $6\,$GHz of the three designs in \cref{fig:floquetdesign} is computed as a function of the loss tangent $\tan\delta$ with all other conditions fixed. Here, we neglect pump attenuation due to the dielectric loss, as it can be compensated by adjusting the circuit parameters accordingly in the adiabatic Floquet scheme. The left and right vertical gray dashed lines correspond to $\tan\delta=3.4\times10^{-3}$ of the SiO$_x$ capacitors in Refs. \cite{macklin_a_2015, tolpygo_deep_2014} and  $\tan\delta=10^{-6}$ of a typical qubit fabrication process \cite{dial_bulk_2016,calusine_analysis_2018}, respectively.  With SiO$_x$ capacitors, the calculated quantum inefficiency of the homogeneous TWPA in Ref. \cite{macklin_a_2015} increases to $\bar{\eta}_{\mathrm{loss}}=0.20$ from its lossless value $\bar{\eta}_{\mathrm{lossless}}= 0.13$, again consistent with the characterized intrinsic quantum inefficiency of $\bar{\eta}_{\mathrm{loss}}=0.25$ in Ref. \cite{macklin_a_2015}. It is worth noting that the presented quantum inefficiency values $\bar{\eta}$ of the Floquet TWPA design are on the same order but not optimal at each loss tangent: for instance, one can leverage the amount of coherent (sideband) and incoherent (dielectric) loss and optimize the net quantum efficiency accordingly with a carefully chosen device length and nonlinearity profile.

The $\bar{\eta}$ of the Floquet scheme rapidly diminishes with a smaller $\tan\delta$ and eventually approaches approximately $10^{-4}$, which is limited by the small impedance mismatch and the finite ramp rate of $I_{pn}(x)$ in \cref{fig:floquetdesign}\textcolor{blue}{(f)}. Floquet mode JTWPAs fabricated with a typical qubit fabrication process are predicted to have a quantum efficiency on the level of $ \eta/\eta_{\mathrm{ideal}}>99.9\%$ ($\bar{\eta}\! <\!10^{-3}$), demonstrating the practicality of our proposed Floquet scheme.

Finally, we discuss the directionality and the prospect of directly integrating a Floquet mode TWPA on chip. In a typical superconducting quantum experiment setup, the preamplifier (JPA or JTWPA) in the measurement chain is only indirectly connected to the device under test via a commercial isolator or circulator to prevent reflections from dephasing the qubits or causing parametric oscillations in the amplifier. Such insertion loss occurring before the preamplifier will degrade the measurement efficiency appreciably and a directional integrated quantum-limited preamplifier is therefore essential for approaching near-perfect full-chain measurement efficiency. While TWPAs are in principle directional, existing TWPAs cannot fulfill this promise due to their non-negligible reflections, as also evidenced in \cref{fig:floquetdesign}\textcolor{blue}{(g)}. It is worth noting that for well-impedance-matched amplifiers, the major obstacle is in fact the nonlinear forward-backward mode coupling, which is properly captured by the off-diagonal block matrices $\mathbb{K}_{12}(x)$ and $\mathbb{K}_{21}(x)$ constituting $\mathbb{K}(x)$ in \cref{eqn:eom}. In Figs. \ref{fig:integration}\textcolor{blue}{(b)} and \ref{fig:integration}\textcolor{blue}{(c)}, we compare the nonlinear impedance and the signal-reflection spectrum of both the conventional homogeneous scheme and the Floquet mode scheme at $\tan\delta=0$. At nonzero loss tangent, backward propagation gets further attenuated and the signal reflection decreases correspondingly. We observe that the signal reflection in the conventional homogeneous scheme is significantly worse than the Floquet scheme even at near-identical and near-ideal impedance matching conditions. In contrast, the Floquet mode TWPA minimizes the nonlinear coupling contribution due to the adiabatic Floquet mode transformation and achieves $<\!-25\,$dB reflection over the entire amplifying bandwidth. To support the claim that the signal reflection of a Floquet mode TWPA is near ideal and limited by impedance mismatch at the boundaries, we simulate the Floquet scheme again using the exact same configurations but manually disabling the nonlinear forward-backward couplings by setting $\mathbb{K}_{12}(x)=\mathbb{K}_{21}(x)=\mathbf{0}_m$, which is plotted in gray in \cref{fig:integration}\textcolor{blue}{(c)}. Indeed, the signal reflection of this ``nonlinearly forward-backward decoupled" hypothetical device is almost identical to the actual Floquet TWPA (red) as expected.

To operate TWPAs as true directional amplifiers and minimize the backaction on devices under test, an additional hurdle to overcome is to minimize the pump reflection and signal reflection at the same time. The pump reflection is largely affected by the mismatch between the port impedance and the pump nonlinear impedance of the TWPA at the boundaries. Floquet mode TWPAs are advantageous in achieving this goal: whereas the strong pump tone sees a different dispersion and nonlinear impedance than the weak signal and idler tones due to self-phase rather than cross-phase modulations in both amplifier designs, the Floquet TWPAs further minimize this discrepancy at the boundaries due to the significantly reduced nonlinearity there. The pump reflection is evaluated to be $S_{11,pump}\approx-48.4\,$dB, using the same parameters in calculating \cref{fig:floquetdesign,fig:integration} (for details, see \cref{app:pumpreflection}).

The improved directionality of Floquet-mode TWPAs, as well as the insensitivity to the out-of-band impedance environment described in \cref{sec:insensitivitytooob}, suggests that Floquet mode TWPAs are still favorable to conventional TWPA designs even when existing fabrication processes ($\tan\delta\approx10^{-3}$) are used and the quantum efficiencies of these amplifier designs are similar.

\section{Insensitivity to out-of-band Impedance Environment} \label{sec:insensitivitytooob}

In this section, we discuss the impact of a nonideal impedance environment on the performance of the proposed Floquet mode TWPAs and conventional TWPAs. In the calculations above, the port impedance is assumed to be $Z_0=50\,\Omega$ over the entire frequency range. In practice however, qubits and TWPAs often see a quite different impedance environment at frequencies higher than $16\,$GHz due to wirebonds, attenuators, connectors, circulators, and other microwave components that are not optimized at those out-of-band frequencies. Alternatively, one might be tempted to intentionally engineer the impedance environment of a conventional TWPA to filter out the higher-frequency sidebands. 

In general, the multimode, quantum input-output theory framework presented in this work can model an arbitrary nonideal impedance environment and its effects on TWPA performance using a frequency-dependent port impedance $Z_0(\omega)$. For both of the specific scenarios discussed above, we can use a simple stepwise impedance model of $Z_0(\omega) = 50\,\Omega$ for $ 0 \leq \abs{\omega/(2\pi)} \leq 16\,$GHz and $Z_0(\omega) = Z_\mathrm{ob}$ otherwise to emulate the behavior of large impedance mismatch outside the target frequency range. Here, we use $Z_{\mathrm{ob}}$ to denote the out-of-band impedance.

\begin{figure}[thp]
\noindent
\includegraphics[width=1\linewidth]{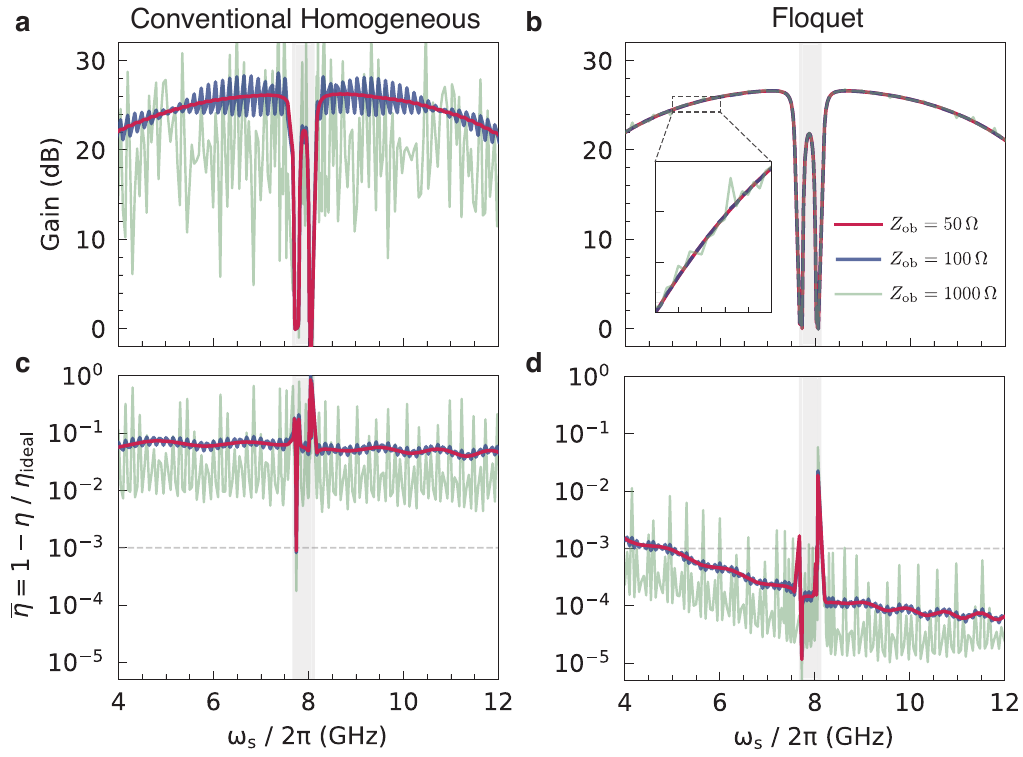}
\caption{The performance of a conventional homogeneous TWPA and a Floquet TWPA with comparable cutoff frequencies under different out-of-band impedance $Z_{\mathrm{ob}} = 50\,\Omega$ (red), $100\,\Omega$ (blue), and $1000\,\Omega$ (green) respectively. The impedance environment is set to be $Z=50\,\Omega$ for $\abs{\omega_s}/(2\pi) \leq 16\,$GHz and $Z=Z_{\mathrm{ob}}$ otherwise. (a),(b) The gain spectra of the conventional TWPA and the Floquet TWPA, respectively. (c), (d) The quantum inefficiency $\bar{\eta}=1-\eta/\eta_{\mathrm{ideal}}$ of the conventional TWPA and the Floquet TWPA respectively. }
\label{fig:nonidealimpedance}
\end{figure}

\Cref{fig:nonidealimpedance} compares the numerically simulated performance of a conventional TWPA and our proposed Floquet TWPA at several out-of band impedance values $Z_\mathrm{ob} = 50\,\Omega$, $100\,\Omega$, and $1000\,\Omega,$ respectively, with all other settings kept equal as described in \cref{app:circuitparams} and \cref{sec:floquettwparesults}. $Z_\mathrm{ob} = 100\,\Omega$ corresponds to an out-of-band linear reflection of approximately $-10\,$dB due to, for example, the high-frequency behavior of wirebonds, whereas $Z_\mathrm{ob} = 1000\,\Omega$ emulates both the typical out-of-band response of commercial isolators and circulators and the intentional filtering of sideband frequencies $>16\,$GHz. As the out-of-band mismatch increases, we see that the gain ripples of the conventional TWPA increase drastically. In the case of $Z_\mathrm{ob} = 1000\,\Omega$, filtering out higher-frequency sidebands does help decrease the quantum inefficiency $\bar{\eta}$ down to approximately $1\%$ at gain-ripple peaks, but at the same time increases $\bar{\eta}$ to as large as approximately $60\%$ at gain-ripple troughs. The large variation in both gain values and quantum efficiency with respect to frequency thus makes this scheme unattractive for applications requiring broad and uniform gain and quantum efficiency.

In stark contrast, the proposed Floquet TWPA is significantly less susceptible to changes in out-of-band impedance environments as evidenced by the minimal changes in its gain profiles. We see that the quantum inefficiency $\bar{\eta}$ remains below $1\%$ at all frequencies and is still superior overall. The drastic difference between the responses of a conventional TWPA and a Floquet TWPA here can be explained by the efficient sideband suppression of Floquet TWPAs.  While the round-trip loss of sidebands decreases significantly under poorly controlled or intentionally engineered low-pass out-of-band impedance environments,
the round-trip gain of a Floquet TWPA remains minimal and significantly smaller than that in a conventional TWPA, making the undesirable parametric oscillations of sidebands much less likely in Floquet TWPAs. This suggests that Floquet TWPAs have the additional practical advantage of being insensitive to the out-of-band impedance environment, which could drastically reduce the design complexity and control requirements of the experimental setup. In addition, it also implies that low-loss Floquet TWPAs can be realistically implemented with distributed capacitors and resonators in a high-quality qubit process with minimal sacrifice in performance (for details, see \cref{app:tlr}). This is because aside from parasitics, such distributed capacitive elements differ slightly from their ideal lumped-element counterparts on dispersion and impedance at high frequencies (see \cref{app:spatialeom}), to which the Floquet TWPAs are shown to be much less insensitive. 

\section{Conclusion}

In conclusion, we propose an adiabatic Floquet mode scheme that allows for both high gain and near-ideal quantum efficiency over a large instantaneous bandwidth. In the cQED platform, we show in calculations that a Floquet mode JTWPA can achieve $>\!20\,$dB gain, $1-\eta/\eta_{\mathrm{ideal}}\!<\!10^{-3}$, and $<\!-20\,$dB reflection over $6.5\,$GHz of instantaneous bandwidth, using a fabrication process with $\tan\delta\sim\!10^{-6}$, typical of qubit fabrication. Crucially, the proposed Floquet mode TWPAs are directional and can thus be directly integrated on chip, potentially leading to near-perfect full-chain measurement efficiency. In addition, their insensitivity to the out-of-band impedance environment, due to sideband suppression, significantly mitigates gain ripples, thus reducing parametric oscillations and instability. We expect this general Floquet mode amplifier paradigm to have far-reaching applications on amplifiers in various platforms and pave the way for scalable fault-tolerant quantum computing.

\section*{Acknowledgements}

This work was funded in part by the AWS Center for Quantum Computing and by the Massachusetts Institute of Technology (MIT) Research Support Committee from the NEC Corporation Fund for Research in Computers and Communications. J.W. acknowledges support from the MIT Center for Quantum Engineering (CQE)-Laboratory for Physical Sciences (LPS) Doc Bedard Fellowship. G.D.C. acknowledges support from the Harvard Graduate School of Arts and Sciences Prize Fellowship. Y.Y. acknowledges support from the National Sciences and Engineering Research Council (NSERC) Postgraduate Scholarship.

K.P.O. and K.P. proposed the adiabatic Floquet mode scheme. K.P. and K.P.O. formulated the multimode quantum input-output theory framework. K.P. and M.N. developed the field ladder-operator-basis model. K.P. and J.W. developed the second-order quantum loss model. K.P. and K.P.O. developed the numerical simulation codes. K.P., M.N., G.D.C., and Y.Y. prepared the figures for the manuscript. K.P. wrote the manuscript with input from all coauthors. K.P.O. supervised the entire scope of the project.\\

\appendix

\section{Circuit Parameters \label{app:circuitparams}}
\Cref{tab:floquetbasisparams} lists the circuit parameters used to calculate \cref{fig:floquetbasis} and the traces corresponding to the conventional TWPA design with $f_c=73$ GHz in \cref{fig:floquetbasis,fig:floquetdesign,fig:integration}.  They are similar to those in Ref. \cite{macklin_a_2015}, except here we assume zero loss or $\tan\delta = 0$. Each unit cell has only one junction. \Cref{tab:floquetdesignparams} lists the circuit parameters used to calculate \cref{fig:floquetdesign,fig:integration,fig:nonidealimpedance}. Each unit cell has two identical junctions in series. In both tables the length of a single unit cell is denoted by a.

\begin{table}[htb]
\caption{\label{tab:floquetbasisparams}Circuit parameters for \Cref{fig:floquetbasis}.}
\begin{ruledtabular}
\begin{tabular}{cccccc}
$I_0 (\mu A)$ & $C_J$(fF)  & $C_g$(fF)  & $C_c$(fF) & $L_r$(pH) \\
\hline
4.55 & 55 & 45 & 20 & 170\\
\hline
 $C_r$(pF) & L (a) & PMR Period (a) & $\omega_s/(2\pi)$ (GHz)  \\
 \hline
 2.82 & 2037 & 3 & 6
 \vspace{-2pt}
\end{tabular}
\end{ruledtabular}
\end{table}

\begin{table}[htb]
\caption{\label{tab:floquetdesignparams}Circuit parameters for the Floquet mode TWPA in \Cref{fig:floquetdesign}.}
\begin{ruledtabular}
\begin{tabular}{cccccc}
$I_0 (\mu A)$ & $C_J$(fF)  & $C_g$(fF)  & $C_c$(fF) & $L_r$(pH) \\
\hline
3.5 & 40 & 76.2 & 40 & 247 \\
\hline
 $C_r$(pF) & L (a) & PMR Period (a) & $\omega_s/(2\pi)$ (GHz)  \\
 \hline
 1.533 & 2000 & 8 & 6 
 \vspace{-2pt}
\end{tabular}
\end{ruledtabular}
\end{table}

\section{MultiMode Quantum Input-Output Theory}

\subsection{Lagrangian and Hamiltonian \label{app:lagrangianhamiltonian}}
\vspace{-10pt}
\begin{figure}[thp]
\centering
\includegraphics[width=0.7\linewidth]{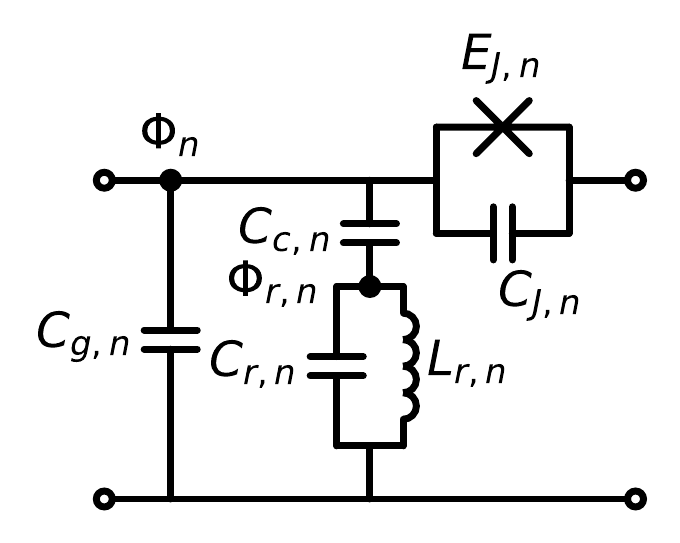}
\caption{Generic unit cell design of a resonantly phase-matched JTWPA. The subscript $n$ denotes the n-th unit cell and its circuit element values can vary from cell to cell.} 
\label{fig:unitcell}
\end{figure}

We consider the unit cell design of a generic resonantly phase-matched Josephson traveling-wave parametric amplifiers (JTWPAs) shown in \cref{fig:unitcell}. It is similar to but more general than those in \cite{obrien_resonant_2014,macklin_a_2015}, as we allow the circuit parameters to have an arbitrary spatial dependence denoted with the subscript $n$ to model Floquet mode TWPAs. The circuit Lagrangian can be expressed as

{
\begin{align}
 &L = \sum_{n}\Bigg[ E_{J, n} \cos \left(\frac{\Phi_{n+1}-\Phi_{n}}{\phi_{0}}\right) +\frac{C_{g, n}}{2}\left(\frac{d \Phi_{n}}{d t}\right)^{2}\nonumber \\
 &\quad +\frac{C_{J, n}}{2}\left(\frac{d \Phi_{n+1}}{d t}-\frac{d \Phi_{n}}{d t}\right)^{2} +\frac{C_{c, n}}{2}\left(\frac{d \Phi_{n}}{d t}-\frac{d \Phi_{r, n}}{d t}\right)^{2} \nonumber \\
 &\quad +\frac{C_{r, n}}{2}\left(\frac{d \Phi_{r, n}}{d t}\right)^{2}-\frac{\Phi_{r, n}^{2}}{2 L_{r, n}}\Bigg] \nonumber \\
 &\approx \int_0^L \frac{1}{a} dx ~~\Bigg(E_J(x)\cos(\frac{\Phi_x a}{\phi_0})+ \frac{C_J(x)}{2}\Phi_{xt}^2 + \frac{C_g(x)} {2}\Phi_t^2 \nonumber \\
 &\quad + \frac{C_c(x) }{2}(\Phi_t - \Phi_{r,t})^2 + \frac{C_r(x) }{2}\Phi_{r,t}^2 -\frac{\Phi_r^2 }{2L_r(x)} \Bigg),
 \label{eqn:origlagrangian}
\end{align}
}

\noindent where in the last step we take the continuum approximation $\Phi_{n+1} - \Phi_{n} \approx a\Phi_{x}(x)\rvert_{x=na}$ and $\sum_n{a} \approx \int{dx}/a$, assuming that the unit cell length $a$ is much smaller than the characteristic wavelength of the system. Here, we also use the subscript notation $\Phi_{x(t)} \equiv \partial{\Phi}/\partial{x(t)} $ to denote partial derivatives with respect to x(t). To simplify notations, we introduce the normalized units and dimensionless variables
{
\begin{equation}
    \begin{aligned}
            &\omega_c = \frac{1}{\sqrt{L_{J0}C_{g0}}}, \ \  \tilde{\phi}=\frac{\Phi}{\phi_{0}}, \ \  \tilde{\psi}=\frac{\Phi_r}{\phi_{0}},\ \  \tilde{\omega}=\frac{\omega}{\omega_{c}}, \\
            & \tilde{t} =\omega_{c}t= t /\tau_{c},  \ \  \tilde{k}=k \cdot a, \ \  \tilde{L} = L/a, \text{~ and} \ \ \tilde{x}=x/a,
    \end{aligned} 
\end{equation}
}
\noindent where $\phi_0 = \Phi_0 / (2\pi)$ is the reduced flux quantum and $L_{J0} = \phi_0^2/E_{J0}$ is the reference junction inductance of choice (For Floquet mode TWPAs, we choose the reference to be at the center where the effective drive amplitude is maximum). The Lagrangian in \cref{eqn:origlagrangian} can now be equivalently expressed with the dimensionless variables $\tilde{t}, \tilde{x}, \tilde{\phi}(\tilde{x},\tilde{t}), \tilde{\psi}(\tilde{x},\tilde{t})$ as

{
\begin{align}
    L &= \int dx\mathcal{L} =  E_{J0}\int_0^{\tilde{L}} dx \Bigg(\mu(x)\Big(\cos(\tilde{\phi}_{\tilde{x}}) + \frac{\beta}{2}\tilde{\phi}_{\tilde{x}\tilde{t}}^2\Big) \nonumber \\
    &+ \frac{\nu(x)}{2}\tilde{\phi}_{\tilde{t}}^2 + \frac{\gamma(x)}{2}(\tilde{\phi}_{\tilde{t}}-\tilde{\psi}_{\tilde{t}})^2 + \frac{\tilde{C}_r}{2}\tilde{\psi}_{\tilde{t}}^2  - \frac{\tilde{\psi}^2}{2\tilde{L}_r}\Bigg) \label{eqn:normlagrangian},
\end{align}
}
\noindent in which the dependence of $\tilde{\phi}, \tilde{\psi}$ on $\tilde{x},\tilde{t}$ is implicitly assumed, and

\begin{align}
    &\mu(x) = \frac{E_J(x)}{E_{J0}} = \frac{L_{J0}}{L_J(x)}= \frac{C_J(x)}{C_{J0}},~ \nu(x) = \frac{C_g(x)}{C_{g0}}, \nonumber \\
    &\beta =\frac{C_{J0}}{C_{g0}}, ~\gamma(x) =\frac{C_{c}(x)}{C_{g0}},  ~\tilde{C}_{r}=\frac{C_{r}}{C_{g0}},  ~\tilde{L}_{r}=\frac{L_{r0}}{L_{J 0}}
\end{align} \label{eqn:scalingcoeffs}

\noindent are the dimensionless parameters describing the spatial profile of the circuit elements. We note that implicit in \cref{eqn:normlagrangian} and \cref{eqn:scalingcoeffs} is the assumption that the plasma frequency $\omega_J(x) = 1 / \sqrt{L_J(x)C_J(x)} = 1/\sqrt{L_{J0}C_{J0}}$ of all the junctions remains constant, as was also assumed in the main text. $C_J(x)$ thus scales with junction critical current $I_0(x)$ and is accounted for by the prefactor $\mu(x)$ of the junction capacitance term $\tilde{\phi}_{\tilde{x}\tilde{t}}^2$ in \cref{eqn:normlagrangian}. This can be conveniently implemented to control the junction properties in experiment by changing the junction area. Unless otherwise noted, we will work entirely in the normalized unit from now on and omit all tildes for brevity.

Following similar procedures in \cite{grimsmo_squeezing_2017}, we identify the dimensionless node fluxes $\phi(x,t)$ and $\psi(x,t)$ as canonical coordinates, and the corresponding canonical momenta $\pi_{\phi}(x,t)$ and $\pi_{\psi}(x,t)$ are therefore

{
    \begin{align}
        \pi_{{\phi}}(x,t) & = \frac{\delta L}{\delta {\phi}_{{t}}} = E_{J0}\bigg\{\nu(x){\phi}_{{t}} - \beta \dv{\left(\mu(x){\phi}_{{x}{t}}\right)}{x} \nonumber \\
        &+ \gamma(x) ({\phi}_{{t}}-{\psi}_{{t}}) \bigg\}, \nonumber \\ 
        \pi_{{\psi}}(x,t) & = \frac{\delta L}{\delta {\psi}_{{t}}}  = E_{J0}(\tilde{C}_r {\psi}_{{t}} + \gamma(x) ({\psi}_{{t}}-{\phi}_{{t}}) ).
    \end{align} 
}
Applying the Legendre transform, we arrive at the Hamiltonian 

{
\begin{align}
    H &=\int_{0}^{L} d x\left( \pi_\phi \phi_{t}+ \pi_\psi \psi_{t}-\mathcal{L} \right) \\
    &=  E_{J0} \int_0^L dx \Big(-\mu(x)\cos(\phi_x) - \phi_{t}\dv{[{\mu(x)\beta}\phi_{xt}]}{dx}  \nonumber \\
    &\quad  -\frac{\mu(x)\beta}{2}\phi_{xt}^2 +  \frac{\nu(x)}{2}\phi_t^2  + \frac{\gamma(x)}{2}(\phi_t-\psi_t)^2 \nonumber \\
    &\quad + \frac{\tilde{C}_r}{2}\psi_t^2 + \frac{\psi^2}{2\tilde{L}_r} \Big)\nonumber  \\
    &= E_{J0} \int_0^L dx \Big(-\mu(x)\cos(\phi_x) + \frac{\mu(x)\beta}{2}\phi_{xt}^2 +  \frac{\nu(x)}{2}\phi_t^2  \nonumber \\
    & \quad  + \frac{\gamma(x)}{2}(\phi_t-\psi_t)^2 + \frac{\tilde{C}_r}{2} \psi_t^2+ \frac{\psi^2}{2\tilde{L}_r} \Big) \nonumber \\
    &\quad +  \text{boundary terms},
\end{align}
}

\noindent where in the last step we perform integration by parts on the term $-\phi_{t}d({\beta}\mu(x)\phi_{xt})/dx$ and produce the additional constant boundary terms, which will be dropped from now on for analyzing dynamics inside the TWPA.

We quantize the system by promoting the variables to operators
\begin{equation}
    \begin{aligned}
        \phi(x,t) &\rightarrow \hat{\phi}(x,t), \quad \psi(x,t) \rightarrow \hat{\psi}(x,t), \\
        \quad \pi_{\phi}(x,t) &\rightarrow \hat{\pi}_{\phi}(x,t), \quad \pi_{\psi}(x,t)  \rightarrow \hat{\pi}_{\psi}(x,t),
    \end{aligned}
\end{equation}
\noindent such that they obey the commutation relations

{
\begin{align}
     \comm{\hat{\psi}(x,t)}{\hat{\pi}_{\phi}(x',t)}&= \comm{\hat{\phi}(x,t)}{\hat{\pi}_{\psi}(x',t)} \nonumber \\
     &= \comm{\hat{\phi}(x,t)}{\hat{\psi}(x',t)}= 0, \nonumber \\
      \comm{\hat{\phi}(x,t)}{\hat{\pi}_{\phi}(x',t)} &= \comm{\hat{\psi}(x,t)}{\hat{\pi}_{\psi}(x',t)}= i\hbar\delta(x-x'). 
\end{align}
}

\subsection{Quantum Spatial Equation of Motion \label{app:spatialeom}}
The quantum spatial equations of motion can thus be derived from the Heisenberg equations of motion:

{
\begin{align}
    & \dv{\hat{\pi}_\phi(x,t)}{t} = \frac{1}{i\hbar}\comm{\hat{\pi}_\phi(x,t)}{H}  \nonumber \\
    & \quad \Rightarrow \dv{x}\left(\mu(x)\left(\sin(\hat{\phi}_x)+\beta\hat{\phi}_{xtt}\right)\right) \nonumber \\
    &\quad = \nu(x)\hat{\phi}_{tt} + \gamma(x)(\hat{\phi}_{tt} - \hat{\psi}_{tt}) \label{eqn:exacteom1}
\end{align} 
\begin{align}
    & \dv{\hat{\pi}_\psi(x,t)}{t} \frac{1}{i\hbar}\comm{\hat{\pi}_\psi(x,t)}{H} \Rightarrow \nonumber \\
    &\qq{} \quad -\frac{\hat{\psi}}{\tilde{L}_r} = \gamma(x)(\hat{\psi}_{tt} - \hat{\phi}_{tt}) + \tilde{C}_r\hat{\psi}_{tt}.\label{eqn:exacteom2}
\end{align} 
}

To make further progress, we make the stiff-pump approximation $\hat{\phi}(x)\rightarrow \phi_p(x) + \hat{\phi}(x)$, in which $\phi_p(x)$ is a classical number that is solved independently from the dynamics of signals and sidebands. Moreover, we neglect the generation of the pump higher
harmonics $3\omega_p$, $5\omega_p$, $\dots$ and solve for the fundamental frequency pump consistently in the form of $\dv*{\phi_p}{x} = A_{px0}(x)\sin(\omega_p t - \int_0^xdx' k_p(x'))$. It should be pointed out that although here we neglect the higher harmonics of the pump, higher order nonlinear processes 4WM, 6WM, $\dots$ from the higher order junction nonlinearities mediated by the fundamental frequency pump are all accounted for and treated appropriately. After performing Fourier transform and cross-eliminating $\hat{\psi}(x,\omega)$, we finally arrive at the single-variable equation of motion in the flux basis

{
\begin{equation}
\begin{aligned}
    & -(\nu(x)+\gamma\alpha_r(\omega))\omega^2\hat{\phi}(\omega) = \dv{x}\Big( -\mu(x)\beta\omega^2 \hat{\phi}_x(\omega) \\
    &+ \mu(x)\sum_{n=-\infty}^\infty J_{2n}(A_{px0}(x)) \hat{\phi}_x(\omega+2n\omega_p)e^{-i2n\int_0^x dx' k_p(x')} \Big),
    \label{eqn:fluxeomfreqbasis}
\end{aligned}
\end{equation}
}
\noindent where $J_{2n}(z)$ is the Bessel function of the first kind of order $2n$, and 
{
\begin{equation}
    \alpha_r(\omega) = \frac{1 - \tilde{L}_r\tilde{C}_r\omega^2}{1 - \tilde{L}_r(\tilde{C}_r+\gamma)\omega^2} = \frac{1-\omega^2/\omega_r^2}{1-\omega^2/\omega_{rt}^2} 
\end{equation}
}
\noindent accounts for the effect of the phase matching resonators (PMRs) and acts as an effective frequency-dependent capacitor. Notice that the cross-elimination is only valid when the frequency $\omega$ is away from the resonance $\omega_{rt}$ \cite{grimsmo_squeezing_2017}. From the left-hand side of \cref{eqn:fluxeomfreqbasis}, we also see that the coupling strength of PMRs is described by $\gamma(x)/\nu(x) = C_c(x)/ C_g(x)$, which can be made constant to maintain a similar phase matching condition across the device.

For an injected signal at frequency $\omega_s = \omega_0$, the only frequency components it can couple to are $\omega_n = \omega_s + 2n\omega_p$, where $n$ is any integer. In practice however, $n$ cannot be an arbitrarily small(negative) or large(positive) due to the restrictions of the junction plasma frequency and the transmission line cutoff frequency (from the discreteness of lumped-element transmission lines).  We can therefore truncate the number of frequency components coupled to the signal to a finite number $m=n_{\mathrm{max}} - n_{\mathrm{min}}+1$ and define a flux operator vector as
\begin{equation}
    \vec{\Phi}(x) = [\cdots, \hat{\phi}(\omega_n) \cdots]^T.
\end{equation}

We can now rewrite \cref{eqn:fluxeomfreqbasis} as a matrix equation in block matrix format:
{
\begin{equation}
    \dv{x} \left(-\mathbb{L}^{-1}(x)\vec{\Phi}_x(x)\right) = -\mathbb{C}(x)\mathbb{W}^2\vec{\Phi}(x), \label{eqn:hermitianodeform}
\end{equation}
}
\noindent in which the normalized $m\times m$ frequency, inductance, and capacitance block matrices are defined as

\begin{widetext}
\begin{align}
    \mathbb{W} &= \mathrm{diag}(\cdots, \omega_n, \cdots), \\
    \mathbb{C}(x) &=  \mathrm{diag}(\cdots, \nu(x) + \gamma(x)\alpha_r(\omega_n), \cdots), \qq{and} \\
    \mathbb{L}^{-1}(x) &= -\mu(x)\beta\mathbb{W}^2 + \mu(x)\begin{pmatrix}
 J_0\theta_0 & J_{-2} \theta_{-2} & J_{-4}\theta_{-4}  & \dots &  J_{2n_{\mathrm{min}}} \theta_{2n_{\mathrm{min}}} \\
 J_2\theta_{2} & J_0\theta_{0} & J_{-2}\theta_{-2} & \ddots &  \vdots \\
J_4\theta_{4} & J_2\theta_{4}  & \ddots & &  \vdots\\
\vdots & \ddots  & &  J_0\theta_{0} & J_{-2}\theta_{-2}\\
J_4\theta_{4} & J_2\theta_{2}  & \ddots & &  \vdots\\
J_{2n_{\mathrm{max}}}\theta_{2n_{\mathrm{max}}}  & \dots  & & J_{2}\theta_{2} & J_0\theta_{0}
\end{pmatrix}.
\end{align}
\end{widetext}

The last term in $\mathbb{L}^{-1}(x)$ is a Toeplitz matrix, and we use the notations $J_{2n} \equiv J_{2n}(A_{px0}(x))$ and $\theta_{2n} =\theta_{2n}(x) \equiv \exp(-i2n\int_0^xdx'k_p(x'))$ for readability purposes. Note that as long as no $\omega_n$ is outside of cutoff frequency or fall in between the resonant bandgap $[\omega_r, \omega_{rt}]$ and the pump current is below the junction critical current, $\mathbb{W}, \mathbb{C}(x),$ and $\mathbb{L}^{-1} (x)$ are all positive-definite matrices, and therefore the inverse of $\mathbb{L}^{-1}(x)$ or $\mathbb{L}(x) \equiv \left(\mathbb{L}^{-1}(x))\right)^{-1}$ exists and is well-defined.

Although here we only consider the case of linear capacitors and PMRs connecting nodes to ground as described by a diagonal $\mathbb{C}(x)$, it is worth noting that our presented input-output quantum framework is general and capable of modeling lossless nonlinear capacitors or any blackbox design described by a diagonal admittance matrix $\mathbb{Y}(x)$, with $\mathbb{Y}_{nn}(x) = Y(\omega_n)$ being the admittance at frequency $\omega_n$. In this case, the corresponding block capacitance matrix becomes $\mathbb{C}(x) = -i\mathbb{W}^{-1}\mathbb{Y}(x)/C_{g0}$. As an example, the diagonal capacitance matrices for a distributed coplanar stub capacitor and a $\lambda/4$ transmission line resonator are $\mathbb{C}_{nn}(x) = \tan(k_{\mathrm{tlr}}l_{\mathrm{tlr}})/(\omega_n C_{g0} Z_{\mathrm{tlr}})$ and $\mathbb{C}_{nn}(x) = -\cot(k_{\mathrm{tlr}}l_{\mathrm{tlr}})/(\omega_n C_{g0} Z_{\mathrm{tlr}})$, respectively. Here, $k_{\mathrm{tlr}}$, $Z_{\mathrm{tlr}}$, and $l_{\mathrm{tlr}}$ are the wavevector, characteristic impedance, and physical length of the transmission line resonators, respectively, and here $\mathbb{W}$ and $\omega_n$ are in the unnormalized frequency unit (rads).

We now define the diagonal nonlinear impedance matrix of the TWPA
{
\begin{equation}
    \mathbb{Z}(x) = \mathrm{diag}(\cdots, Z_n(x), \cdots) = \mathrm{diag}(\cdots, \sqrt{\frac{\mathbb{L}_{nn}}{\mathbb{C}_{nn}}}, \cdots), 
\end{equation}
}
\noindent where $\mathbb{L}_{nn}$ and $\mathbb{C}_{nn}$ denote the n-th diagonal elements of the two matrices. As can be observed later in the boundary condition calculations, the diagonal element $\mathbb{Z}_{nn}(x)$ indeed represents the effective nonlinear impedance of mode $n$. Finally, applying the transformation

{
\begin{equation}
    \begin{aligned}
        \vec{\phi}(x) &\rightarrow i \sqrt{\frac{2\phi_0}{\hbar}} \mathbb{W}^{-1}\mathbb{Z}^{1/2}(x)\abs{\mathbb{W}}^{-1/2}(\vec{A}^+(x) + \vec{A}^-(x)) \\
        \vec{\phi}_x(x) &\rightarrow -\sqrt{\frac{2\phi_0}{\hbar}}\mathbb{L}(x)\mathbb{Z}^{-1/2}(x)\abs{\mathbb{W}}^{-1/2}(\vec{A}^+(x) - \vec{A}^-(x)),
    \end{aligned} \label{eqn:flux2laddersm}
\end{equation}
}
we arrive at the field ladder operator basis equation of motion
{
\begin{align}
    \dv{x} \begin{pmatrix} 
 \vec{A}^+ (x)\\  \vec{A}^-(x)
\end{pmatrix} &=  \mathbb{K}(x) \begin{pmatrix}  \vec{A}^+ (x)\\  \vec{A}^-(x)
\end{pmatrix} \nonumber \\ 
&= \begin{pmatrix} 
 \mathbb{K}_{11}(x) &  \mathbb{K}_{12}(x) \\  \mathbb{K}_{21}(x) &  \mathbb{K}_{22}(x)
 \end{pmatrix}\begin{pmatrix} 
 \vec{A}^+ (x)\\  \vec{A}^-(x)
\end{pmatrix},
\end{align} \label{eqn:losslesseom}  
}
\noindent in which

\begin{widetext}
\begin{equation}
    \begin{aligned}
        \mathbb{K}_{12}(x) &= \mathbb{K}_{21}^\dagger(x) = \mathbb{Z}_x(x) \mathbb{Z}^{-1}(x) +\frac{i}{2}\bigg(\abs*{\mathbb{W}}^{-1/2}\mathbb{W}  \big(-\mathbb{Z}^{-1/2}\mathbb{L} \mathbb{Z}^{-1/2} + \mathbb{Z}^{1/2}\mathbb{C}\mathbb{Z}^{1/2} \big)\mathbb{W}\bigg), \\
        \mathbb{K}_{11}(x) &=-\mathbb{K}_{22}(x) = \frac{i}{2}\bigg(\abs*{\mathbb{W}}^{-1/2}\mathbb{W}  \big(\mathbb{Z}^{-1/2}\mathbb{L} \mathbb{Z}^{-1/2}  + \mathbb{Z}^{1/2}\mathbb{C}\mathbb{Z}^{1/2} \big) \mathbb{W}\bigg).
    \end{aligned} \label{eqn:flux2ladder}
\end{equation}
\end{widetext}

\noindent are the $m\times m$ multimode coupling matrices that describe the forward-forward, backward-backward, forward-backward, and backward-forward interactions.

Notice that here we did not apply the usual slowly-varying envelope approximation (SVEA) to reduce the equation of motion in the flux basis to the first order. Going beyond the SVEA allows us to capture the interactions between the forward and backward modes, model the reflection due to impedance mismatch at the boundaries, and crucially to conserve the bosonic commutation relations without making additional \textit{ad hoc} approximations, such as in Ref. \cite{yaakobi_parametric_2013}.

\section{Boundary Conditions and Input-Output Theory \label{app:inputputtheory}}
Assuming the linear transmission lines ports at $x=0$ and $x=L$ to be semi-infinite and have inductance and capacitance per unit length of $l_l$ and $c_l$, we can write the Lagrangian of the extended system as
 
 {
 \begin{equation}
     \begin{aligned}
        L_{\mathrm{full}} &= E_{J0}\int_0^L dx \bigg(\mu(x)\cos(\phi_x) + \frac{\beta}{2}\phi_{xt}^2 + \frac{\nu(x)}{2}\phi_t^2 \\
        &\quad + \frac{\gamma(x)}{2}(\phi_t-\psi_t)^2 + \frac{\tilde{Cr}}{2}\psi_t^2  - \frac{\psi^2}{2\tilde{L_r}}\bigg)\\
        & \quad + \int_{-\infty}^{0}dx \left(\frac{\phi_x^2}{2\tilde{l}_{l}} + \frac{\tilde{c}_{l}\phi_t^2}{2}\right) + \int^{+\infty}_{L}dx \left(\frac{\phi_x^2}{2\tilde{l}_{l}} + \frac{\tilde{c}_{l}\phi_t^2}{2}\right), \label{eqn:discretelagrangian}
    \end{aligned}
 \end{equation}}
 
 \noindent in which $\tilde{l}_l = (l_l \cdot a) / (\phi_0^2 / E_{J0})$ and $\tilde{c}_l = (c_l \cdot a)/ C_{g0}$ are the dimensionless inductance and capacitance parameters of the transmission line ports, and the extended Lagrangian $L_{\mathrm{full}}$ is piece-wise smooth. The continuity of flux $\phi(x)$ and the Lagrange's equations at the boundaries $x=0$ and $x=L$ yield the boundary conditions
 
 {
 \begin{equation}
     \begin{aligned}
        \phi(x=0^-) &= \phi(x=0^+), \\
        \phi(x=L^-) &= \phi(x=L^+), \\
        \phi_x(x=0^-) / \tilde{l}_l &= \beta\phi_{xtt}(x=0^+) + \mu(x)\sin\left(\phi_x(x=0^+)\right), \\ 
        \phi_x(x=L^+) / \tilde{l}_l &= \beta\phi_{xtt}(x=L^-) + \mu(x)\sin\left(\phi_x(x=L^-)\right),
     \end{aligned}\label{eqn:leftbc}
 \end{equation}
 }
 
 \noindent which can be interpreted as the flux (voltage) and current continuity conditions at the boundaries. Performing the stiff-pump-approximation, going into the frequency basis and applying again the transformations in \cref{eqn:flux2laddersm}, we obtain the linearized ladder operator boundary conditions
 
 {
 \begin{equation}
     \begin{aligned}
        \begin{pmatrix}
       \vec{A}^+(0^+) \\ \vec{A}^-(0^+)
     \end{pmatrix} &= \begin{pmatrix}
        \mathbb{BC}_{11}(0^+) &  \mathbb{BC}_{12}(0^+) \\ \mathbb{BC}_{21}(0^+) &  \mathbb{BC}_{22}(0^+)
     \end{pmatrix}\begin{pmatrix}
       \vec{A}^+(0^-) \\ \vec{A}^-(0^-)
     \end{pmatrix} \\ 
     \begin{pmatrix}
       \vec{A}^+(L^-) \\ \vec{A}^-(L^-)
     \end{pmatrix} &= 
     \begin{pmatrix}
        \mathbb{BC}_{11}(L^-) &  \mathbb{BC}_{12}(L^-) \\ \mathbb{BC}_{21}(L^-) &  \mathbb{BC}_{22}(L^-)
     \end{pmatrix}\begin{pmatrix}
       \vec{A}^+(L^+) \\ \vec{A}^-(L^+)
     \end{pmatrix}
     \end{aligned}, \label{eqn:ladderbc}
 \end{equation}}
 
\noindent where the diagonal and off-diagonal matrices are

{
\begin{equation}
\begin{aligned}
   \mathbb{BC}_{11}(x) &= \mathbb{BC}_{22}(x) = \frac{1}{2}\left(\sqrt{\frac{Z_0}{\mathbb{Z}(x)}} + \sqrt{\frac{\mathbb{Z}(x)}{Z_0}}\right) \\ 
   \mathbb{BC}_{12}(x) &= \mathbb{BC}_{21}(x) = \frac{1}{2}\left(\sqrt{\frac{Z_0}{\mathbb{Z}(x)}} - \sqrt{\frac{\mathbb{Z}(x)}{Z_0}}\right),
\end{aligned}
\end{equation}}

\noindent in which $Z_0=\sqrt{l_l/c_l}$ is the characteristic impedance of the input/output transmission line, and we use the notation $\sqrt{\mathbb{Z}(x)} = \mathbb{Z}^{1/2}(x)$ and $1/\sqrt{\mathbb{Z}(x)} = \mathbb{Z}^{-1/2}(x)$.

To formulate the input-output theory, we denote the input and output operator vectors as

\begin{equation}
    \vec{A}_{\mathrm{in}} = 
    \begin{pmatrix}
      \vec{A}_{\mathrm{in}}^+ \\ \vec{A}_{\mathrm{in}}^- 
    \end{pmatrix} = 
    \begin{pmatrix}
       \cdots \\
      \hat{a}^{+}_{n}(0^-) \\
      \cdots \\ \cdots \\
      \hat{a}^{-}_{n}(L^+) \\\cdots
    \end{pmatrix}, \vec{A}_{\mathrm{out}} = 
    \begin{pmatrix}
      \vec{A}_{\mathrm{out}}^+ \\ \vec{A}_{\mathrm{out}}^- 
    \end{pmatrix} = 
    \begin{pmatrix}
       \cdots \\
      \hat{a}^{+}_{n}(L^+) \\
      \cdots \\ \cdots \\
      \hat{a}^{-}_{n}(0^-) \\\cdots
    \end{pmatrix}. \label{eqn:inoutvecs}
\end{equation}

Equations \eqref{eqn:eom} and \eqref{eqn:ladderbc} constitute a two-point boundary value problem and can therefore be numerically solved to obtain the input-output relation in \cref{eqn:scatteringrelation}, with $\mathbb{S}_0$ being the solution to the sourceless system(i.e., \cref{eqn:losslesseom} without the last term $\vec{\mathrm{F}}^\pm(x)$) and $\mathbb{S}_n(x)$ being the Green's matrix solution to the system driven by a single point source $\vec{\mathrm{F}}^\pm(x)$ and satisfying the boundary conditions \cref{eqn:ladderbc}. In the lossless model where $S_n(x) = \mathrm{O}_{2m}$ is zero, the numerically solved $S_0$ preserves the bosonic commutation relations as expected. One can also check that in the full loss model, the numerically solved scattering matrices $\mathbb{S}_0$ and $\mathbb{S}_n(x)$ together preserve the bosonic commutation relations at the output. 

\section{Quantum Loss Model \label{app:lossmodel}}
As illustrated in \cref{fig:setup}\textcolor{blue}{(a)}, dielectric losses can be modeled quantum-mechanically using a series of lossless transmission line ports, whose frequency-dependent scattering parameters are determined by the loss rate $\mathrm{\Gamma}(x)$. Similar to the time-domain Langevin equations, the effect of dissipation and its associated fluctuation on both the forward and backward waves can thus be incorporated into the lossless spatial equation of motion \cref{eqn:losslesseom} to get \cref{eqn:eom} in the main text. The phase factors in front of $\hat{F}^{\pm}(x)$ are arbitrary and are chosen to be 1 for convenience \cite{jeffers_quantum_1993}, as they do not affect the quantum statistics of the outputs.

\section{Floquet mode TWPAs} \label{app:floquet}

\subsection{Floquet Theory}
In the case of a homogeneous TWPA driven with a constant pump, $A_{px0}(x) = A_{px0}$ and $\theta_{2n}(x) = \theta_{2n} = \exp(-i2n k_p x)$. Therefore, the multimode coupling matrix is periodic and has a period of $x_T = \pi/k_p$. We can therefore apply the Floquet theory to analyze the system. Denote the unique frequency-basis transfer matrix solution of \cref{eqn:losslesseom} to be $\Pi(x)$, such that

\begin{equation}
\vec{A} (x) = 
    \begin{pmatrix} 
 \vec{A}^+ (x)\\  \vec{A}^-(x)
\end{pmatrix} = \Pi(x) \begin{pmatrix} 
 \vec{A}^+ (0)\\  \vec{A}^-(0)
\end{pmatrix} = \Pi(x) \vec{A} (0), \label{eqn:tfmatdef}
\end{equation}

\noindent which is an initial value problem and can be solved numerically. The Floquet theorem states that the $\Pi(x)$ can be written in the form of \cite{teschl_ordinary_2012}

\begin{equation}
    \Pi(x) = \mathbb{P}(x)~ \mathrm{exp}(x \mathbb{Q}),
\end{equation}

\noindent where the $2m\times2m$ matrix $\mathbb{P}(x)$ has the same period $x_T$ as $\mathbb{K}(x)$, $\mathbb{P}(0) = \mathbb{I}$ is the identity matrix, and $\mathbb{Q}$ is a constant $2m\times2m$ matrix that can be obtained from the monodromy matrix

\begin{equation}
    \mathbb{M}_0 = \Pi(x_T) = \mathrm{exp}(x_T \mathbb{Q}).
\end{equation}

Applying the transformation $\vec{B} (x) = \mathbb{P}^{-1}(x)\vec{A}(x)$, \cref{eqn:losslesseom} can be now rewritten in the form of a constant dynamic matrix
\begin{equation}
    \dv{x}\vec{B}(x) = \mathbb{Q} \vec{B}(x).
\end{equation}

With the eigendecomposition of $\mathbb{Q}$ to be $\mathbb{Q} = \mathbb{V}\Lambda\mathbb{V}^{-1}$, where $\Lambda = \mathrm{diag}(\cdots,r_{\alpha}, \cdots)$, and the columns of $\mathbb{V}$ are the corresponding set of normalized eigenvectors, we can therefore transform from the frequency basis $\vec{A}(x)$ into the Floquet basis $\vec{Q}(x)$ using

\begin{equation}
    \vec{Q}(x) = \mathbb{V}^{-1}\mathbb{P}^{-1}(x) \vec{A}(x). \label{eqn:floquetbasistransform}
\end{equation}

\begin{figure}[h]
\centering
\includegraphics[width=0.98 \linewidth]{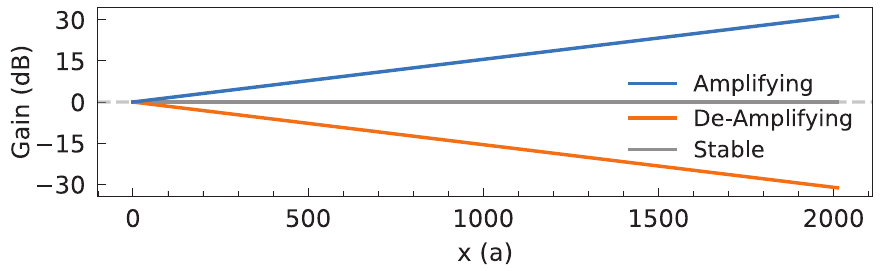}
\caption{Spatial dynamics of Floquet modes in conventional TWPAs. Each curve describes a individual system response in the Floquet basis when a single amplifying ($\hat{Q}_a$, blue), deamplifying ($\hat{Q}_a$, red), or any of the stable (gray) Floquet mode is injected at $x=0$ respectively.} 
\label{fig:floquetdynamics}
\end{figure}

We can gain insights of \cref{eqn:floquetbasistransform} by applying \cref{eqn:tfmatdef} to it:
{
\begin{align}
    \vec{Q}(x) &= \mathbb{V}^{-1}\mathbb{P}^{-1}(x) \vec{A}(x) = \mathbb{V}^{-1}\mathbb{P}^{-1}(x) \Pi(x) \vec{A}(0)\nonumber \\
    & = \mathbb{V}^{-1}\mathbb{P}^{-1}(x) \big(\mathbb{P}(x)\mathrm{exp}(x \mathbb{Q})\big) \vec{A}(0) \nonumber \\
    &= \mathbb{V}^{-1}\big(\mathbb{V}\exp(x\Lambda)\mathbb{V}^{-1}\big)\vec{A}(0) \nonumber \\
    &= \exp(x \Lambda) \big(\mathbb{I}~\mathbb{V}^{-1}\vec{A}(0)\big)\nonumber \\
    &=  \exp(x \Lambda) \big(\mathbb{P}(0)~\mathbb{V}^{-1}\vec{A}(0)\big) = \exp(x \Lambda) \vec{Q}(0),
\end{align}
}
\noindent which shows that Floquet modes $\vec{Q}(x)$ are decoupled from each other and each propagates with a distinct dynamic factor $r_\alpha$, which are also often referred to as the Floquet characteristic exponents. \Cref{fig:floquetdynamics} shows the spatial dynamics of the system in the Floquet basis, with each curve representing the case when a specific Floquet modes is injected at $x=0$. As expected, when only a single Floquet mode is injected, it does not generate or couple to other Floquet modes.

\subsection{Frequency Decomposition of Floquet modes}

\begin{figure}[thp]
\centering
\includegraphics[width=0.98 \linewidth]{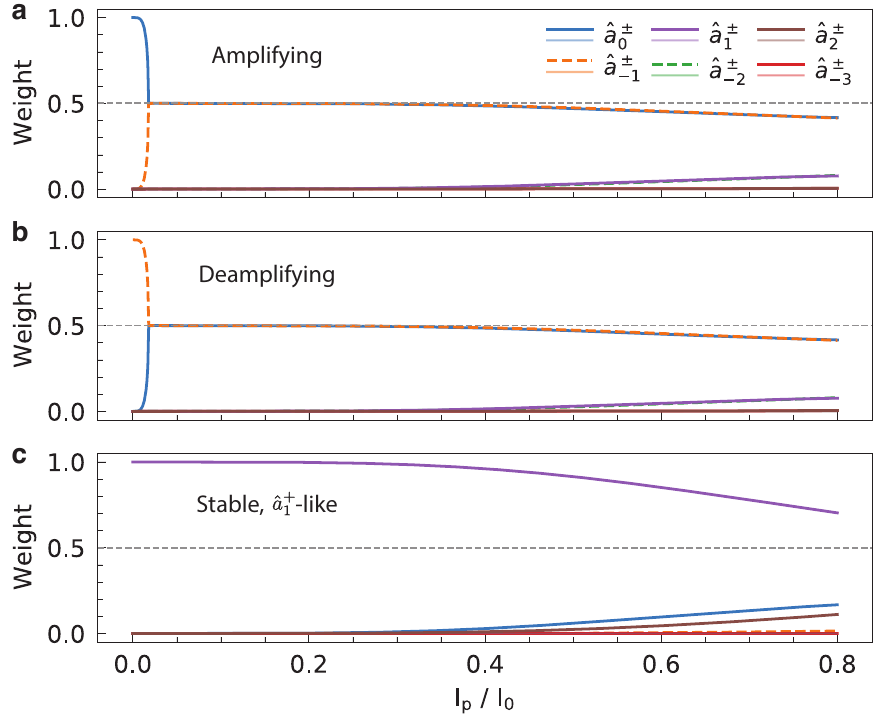}
\caption{Frequency mode decomposition of Floquet modes as a function of pump current. (a) Amplifying Floquet mode $\hat{Q}_a$. (b) Deamplifying Floquet mode $\hat{Q}_d$. (c) A stable Floquet mode that is $\hat{a}^+_{1}$-like.} 
\label{fig:floquetdecomposition}
\end{figure}

We can now analyze the Floquet modes using \cref{eqn:floquetbasistransform}. In \cref{fig:floquetdecomposition} we plot the frequency mode decomposition of three Floquet modes as a function of the dimensionless drive amplitude $I_{pn} = I_p / I_0$. \Cref{fig:floquetdecomposition}\textcolor{blue}{(a)} shows the decomposition of the amplifying Floquet mode $\hat{Q}_a$, which is the same as Fig. 3d in the main text. From the decomposition of the deamplifying Floquet mode $\hat{Q}_d$ in \cref{fig:floquetdecomposition}\textcolor{blue}{(b)}, we see that passing the bifurcation point $I_{pn}\approx 0.02$ the magnitude of the frequency mode decomposition for $\hat{Q}_a$ and $\hat{Q}_d$ are exactly the same and only differ in the relative phase between the frequency components. Therefore, $\hat{Q}_a$ and $\hat{Q}_d$ can be understood as the squeezing and antisqueezing quadratures mostly composed of the signal and idlers.  \Cref{fig:floquetdecomposition}\textcolor{blue}{(c)} corresponds to the frequency mode decomposition of a stable Floquet mode which is $\hat{a}^+_{1}$-like. At high $I_{pn}$, more signal and idler components are mixed in as expected.

\subsection{Gain Scaling} \label{app:gainscaling}
\begin{figure}[thp]
\centering
\includegraphics[width=0.98 \linewidth]{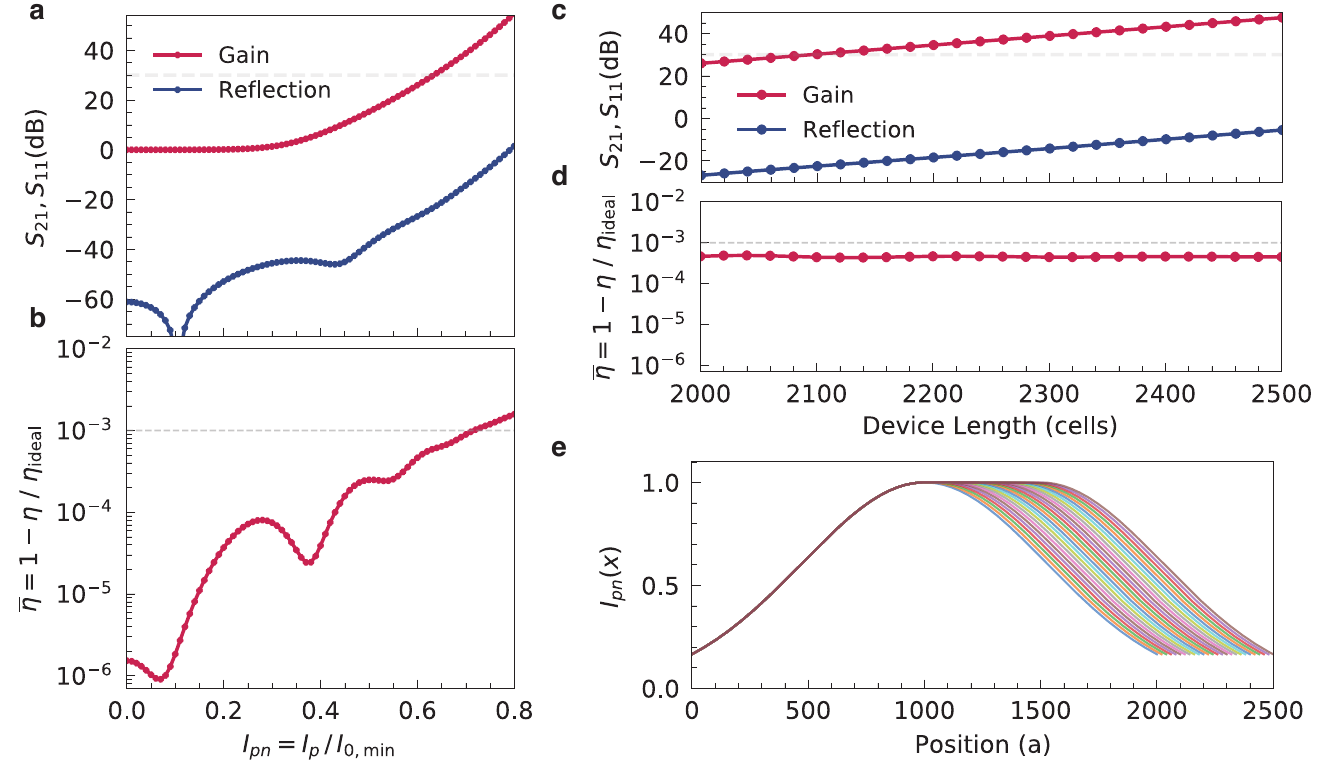}
\caption{Gain scaling of Floquet TWPA at $6\,$GHz with effective pump strength and device length. (a) Gain ($S_{21}$) and reflection ($S_{11}$) with respect to driving strength $I_{pn} = I_p / (I_{0\mathrm{min}})$. (b) Quantum inefficiency  $\bar{\eta} = 1 - \eta/\eta_{\mathrm{ideal}}$ with respect to $I_{pn}$. (c) Gain ($S_{21}$) and reflection ($S_{11}$) with respect to device length. (d) Quantum inefficiency $\bar{\eta}= 1 - \eta/\eta_{\mathrm{ideal}}$ with respect to device length. (c) Profiles of effective pump strength $I_{pn}(x) = I_p / I_0(x)$ used to generate (c),(d).} 
\label{fig:gainscaling}
\end{figure}

In this section, we provide further details about gain scaling on Floquet mode TWPAs. The broadband signal gain of Floquet mode TWPAs can be similarly extended either by increasing the effective pump strength $I_{pn} = I_p / I_{0,\mathrm{min}}$ or with a longer device length. In \cref{fig:gainscaling} we plot the performance scaling of a Floquet TWPA at $6\,$GHz with respect to driving pump strength and device length, respectively. We note that in both scenarios the signal reflection scales with signal gain due to finite reflection at boundaries. Similar to what was described in \cref{sec:floquetpicture} on conventional TWPAs, adjusting the effective pump strength allows the signal gain of Floquet mode TWPAs to be increased in situ at the cost of larger quantum inefficiency $\bar{\eta}$, except such increase is much less pronounced for Floquet TWPAs. This is not surprising, as this degradation in quantum efficiency results from an increase in effective pump strength and therefore sideband coupling, which Floquet TWPAs efficiently suppress.

Increasing signal gain with a slightly longer device length, on the other hand, provides the advantage that the quantum efficiency remains similar at higher gain. Furthermore, the increase in device length to reach a higher gain is modest: this is achieved by modifying the nonlinearity profile and slightly extending the center region near which the gain coefficient is largest (e.g. see \cref{fig:floquetbasis}\textcolor{blue}{(e)}). For instance, using the same parameters in the main text, $\geq30\,$dB gain can be achieved either by driving the same Floquet TWPA design slightly harder at $I_{pn}\sim0.635$ ($6\%$ increase), or by a longer device with $2,100$ cells ($5\%$ increase).

\subsection{Pump Reflection} \label{app:pumpreflection}

Here, we provide further details on the calculation of pump reflection discussed in \cref{sec:directionality}. In line with the stated assumptions in \cref{app:spatialeom}, we express the first order derivative of pump flux in the sinusoidal form of $\phi_{p,x}(x) = A_{px0}(x)\sin(\omega_p t - \int_0^xdx' k_p(x'))$. In Floquet mode TWPAs, the higher pump harmonics $3\omega_p, 5\omega_p, \dots$ can be neglected to a very good approximation, because pump harmonic generation is minimal at the start of Floquet mode TWPAs where nonlinearity is minimal, and the pump is adiabatically transformed through the center region where the nonlinearity is large. Considering only the fundamental pump frequency $\omega_p$, substituting in the expression of $\phi_{p,x}(x)$, performing Fourier transform on the equations of motion \cref{eqn:exacteom1,eqn:exacteom2}, and taking derivative with respect to x on both sides, we arrive at

\begin{equation}
\begin{aligned}
   &\dv[2]{x}(\mu(x)\left(2J_1(A_{px0}(x)) - \beta\omega_p^2 A_{px0}(x)\right)e^{i\int_x k_p(x')dx'}) = \\ &-\omega^2 c_p(x) A_{px0}(x)e^{i\int_0^x k_p(x')dx'}.
 \end{aligned} \label{eq:pumpeom}
\end{equation}

\noindent in which we define the effective pump capacitance as $c_p(x) \equiv c(\omega_p)= \nu(x)+\gamma\alpha_r(\omega_p) $. Because the spatial variation of nonlinearity and dispersion are slow and adiabatic in Floquet TWPAs, we make the approximations $\mu_x(x) / k_p(x),~ \mu_{xx}(x)/ k_p(x)^2 \ll \mu(x)$, and  $A_{px0, x}(x)/k_p(x),A_{px0, xx}(x)/k_p(x)^2 \ll A_{px0}(x)$ analogous to \cite{obrien_resonant_2014}. Expanding out the derivatives in \cref{eq:pumpeom} and applying the above approximations, we arrive at

\begin{equation}
k_p(x) \approx \omega_p \sqrt{\frac{c_p(x)}{\mu(x) \left(\frac{2J_1(A_{px0}(x))}{A_{px0}(x)}- \beta\omega_p^2 \right)}}. \label{eq:kpexpr}
\end{equation}

From \cref{eq:pumpeom,eq:kpexpr}, we see that the effective inductance and capacitance (normalized by $L_J$ and $C_{g0}$ respectively) the pump tone sees are $1/\left(\mu(x) (2J_1(A_{px0}(x))/A_{px0}(x)- \beta\omega_p^2)\right)$ and $c_p(x)$ respectively. We also note that with Bessel function expanded to third order, $2J_1(A_{px0}(x))/A_{px0}(x) \approx 1 - A_{px0}(x)^2/8$ and \cref{eq:kpexpr} recovers the usual self-phase-modulation expression of \cite{obrien_resonant_2014} in which higher than fourth-order junction potential terms are ignored. The nonlinear impedance the pump tone experiences is thus
\begin{equation}
    Z_{p,nl}(x) \approx \sqrt{\frac{ L_J/C_{g0}  }{c_p(x)\mu(x) \left(\frac{2J_1(A_{px0}(x))}{A_{px0}(x)}- \beta\omega_p^2\right)}}.
\end{equation}

Furthermore, within the adiabatic approximation, $\phi_{p,x}$ can be related to the effective pump strength $I_{pn}(x)$ as

{
\begin{equation}
 I_{pn}(x) \approx \frac{I_{p,in}}{I_0(x)} = \frac{I_p}{\mu(x)I_0}= 2J_1(A_{px0}(x)) - \beta\omega_p^2 A_{px0}(x),
\label{eqn:phicurrentconversion}
\end{equation}
}

\noindent where $I_0(x)$ is the junction critical current at $x$, and $I_{p,in}$ is the physical input pump current (again neglecting coupling to the higher harmonics of pump). The dimensionless pump flux amplitude $A_{px0}(x)$ can thus be numerically solved from the effective pump strength $I_{pn}(x)$ using \cref{eqn:phicurrentconversion}. Finally, defining the interface reflection coefficients $r_{01,p} = (Z_{p,nl}(0)-Z_0)/(Z_{p,nl}(0)+Z_0)$ and  $r_{12,p} = (Z_0-Z_{p,nl}(L))/(Z_0+Z_{p,nl}(L))$, we can thus estimate pump reflection in the Floquet TWPA by
{
\begin{align}
    \abs{S_{11,pump}} &= \abs{\frac{r_{01,p} + r_{12,p}e^{2i\int_0^L k_p(x)dx}}{1 + r_{01,p} r_{12,p}e^{2i\int_0^L k_p(x)dx}}}^2  \nonumber \\
    &\leq \frac{\abs{ \abs{r_{01,p}} + \abs{r_{12,p}} }^2}{\abs{1 + \abs{r_{01,p}r_{12,p}}}^2}.
\end{align}
}

Using the same parameters as in \cref{fig:floquetdesign,fig:integration,app:circuitparams}, the pump reflection at $7.875\,$GHz is evaluated to be $\sim\!-48.4\,$dB.

\subsection{Performance with Distributed Phase Matching Resonators} \label{app:tlr}

\begin{figure}[bh]
\centering
\includegraphics[width=0.98 \linewidth]{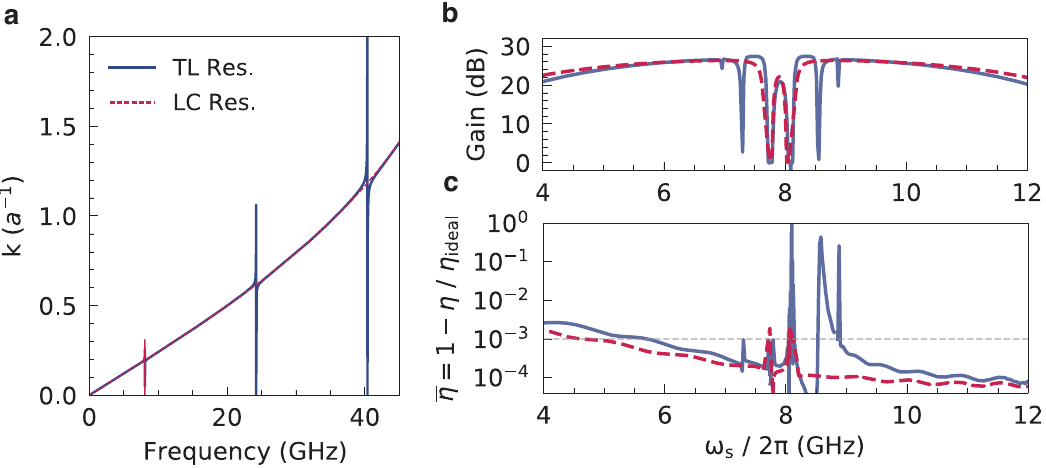}
\caption{Performance of Floquet TWPAs implemented with lumped LC resonators and distributed transmission line resonators (TLRs). The TLRs are assumed to have typical phase velocity of $1.3\times10^8\,$m/s, and their characteristic impedance vary from $50\,\Omega$ to $\sim\!300\,\Omega$ for numerical evaluation convenience. The parameters of the Floquet TWPA with lumped LC resonators are the same as those used in Fig. 4, 5 of the main text, and the parameters of the Floquet TWPA with TLRs are mostly identical with modified $C_c(x)$ and resonator spacing to match the resonator coupling strength. (a) Wavevector (k). (b) Signal gain ($S_{21}$). (c) Quantum inefficiency $\bar{\eta}=  1 - \eta/\eta_{\mathrm{ideal}}$.} 
\label{fig:tlr}
\end{figure}

Heretofore we have assumed ideal lumped LC phase matching resonators in analysis. We now discuss the alternative of employing distributed transmission line resonators (TLRs) \cite{white_traveling_2015} and its effect on the performance of Floquet mode TWPAs. Here we choose to implement the distributed phase matching resonators using a coupling capacitor $C_c(x)$ and $\lambda/4$ TLRs shorted to ground on the other end, because at low frequencies they can be approximated as parallel LC resonators. Following the procedure presented in \cref{app:spatialeom}, we can write the diagonal elements of the resulting capacitance matrix (including also the parallel ground capacitance $C_g(x)$) as 

{
\begin{equation}
\begin{aligned}
    \mathbb{C}_{nn}(x) &= \nu(x) + \frac{1}{i\omega_nC_{g0}\left(\frac{1}{i\omega_n C_c(x)} + iZ_{tlr}\tan(k_{tlr}l_{tlr})\right)} \\ 
    &=  \nu(x) + \frac{\gamma(x)}{1 -\omega_n C_c(x)Z_{tlr}(x)\tan(k_{tlr}l_{tlr})},
\end{aligned}   
\end{equation}
}

\noindent in which the definitions follow those in \cref{app:lagrangianhamiltonian,app:spatialeom}, except here $\omega_n$ is in the unnormalized frequency unit (rads). In \cref{fig:tlr} we compare the numerically simulated performance of Floquet TWPAs implemented with lumped element LC resonators and with distributed $\lambda/4$ transmission line resonators (TLRs), respectively. The parameters of the Floquet TWPA with LC resonators are the same as those used in \cref{fig:floquetdesign,fig:integration}, and the parameters of the Floquet TWPA with TLRs are almost identical except with reduced spacing=1 and $\max[C_c(x)] = C_{c0} =5\,$fF to match the coupling strength of both resonators. Moreover, we assumed a typical phase velocity $v_{ph,tlr}=\omega/k_{tlr}(\omega)=1.3\times10^8\,$m/s for TLRs, and for numerical evaluation convenience we vary the characteristic impedance $Z_{tlr}(x)$ between $50\,\Omega$ and $\sim\!300\,\Omega$ such that $Cc(x)Z_{tlr}(x)$ remains constant.

\begin{figure}[thp]
\centering
\includegraphics[width=0.98 \linewidth]{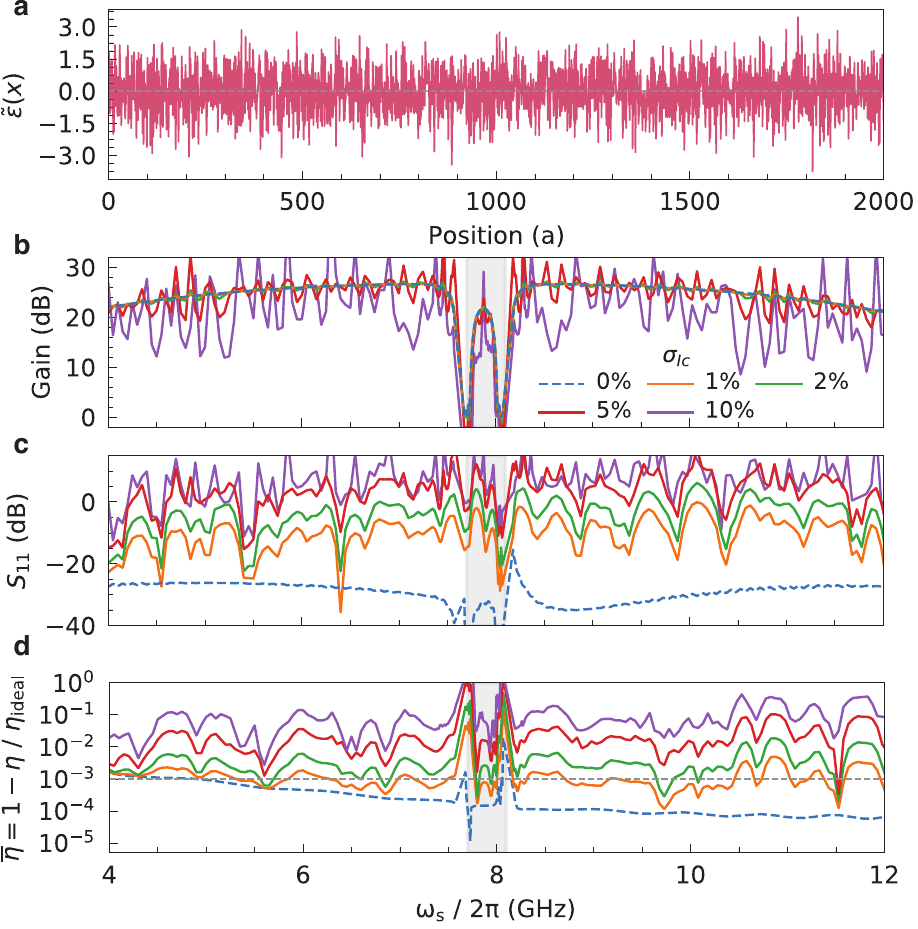}
\caption{Typical performance of the Floquet TWPA design in \cref{app:circuitparams} and \cref{sec:floquettwparesults} under different degrees of junction critical current variation $\sigma_{Ic}$. Subfigures (b), (c), and (d) share the same legend in (b). (a) Random variable profile $\tilde{\epsilon}(x)$ used to generate (b)-(d). (b) Gain. (c) Signal Reflection. (d) Quantum inefficiency $\bar{\eta}$.} 
\label{fig:paramvar}
\end{figure}

As expected, the distributed nature of TLRs results in different dispersion and impedance of sidebands at higher frequencies. We observe that the gain and quantum efficiency are similar over most of the band.  The additional features near $\omega_r$ for the Floquet TWPA with TLRs are due to the modified dispersion of the sideband near $3\omega_r$ phase matching the corresponding sideband coupling process. The weak dependence of performance on resonator implementation details and higher frequency dispersion showcases the robustness and practicality of our proposed Floquet mode TWPA design and makes experimental realization using a low-loss qubit fabrication process feasible.

\subsection{Parameter Variation} \label{app:paramvar}

We now discuss the effects of non-ideal parameter variations to the Floquet TWPA performance. Specifically, we consider the case of spatial junction critical current variation that could result from fabrication non-uniformity. We model junction variation with an effective critical current profile $\tilde{I_0}(x) \equiv I_0(x)(1+\sigma_{Ic}\tilde{\epsilon}(x))$, where $I_0(x)$ is the ideal critical current defined in \cref{sec:floquettwparesults}, $\sigma_{Ic}$ is the standard deviation, and $\tilde{\epsilon}(x)$ is a continuous Gaussian normal random variable with mean $\ev{\tilde{\epsilon}(x)}=0$ and variance $\ev{\Delta\tilde{\epsilon}(x)}^2=1$.

\Cref{fig:paramvar} shows the typical performance of the Floquet TWPA design in the main text under junction variation $\sigma_{Ic}=0\%$, $2\%$, $5\%$, and $10\%$ respectively. The junction critical current variation primarily impacts the reflection, $S_{11}$, and the drastic increase is caused by the random mismatch and linear reflections between adjacent unit cells.
Whereas the directionality objective puts stringent requirements of sub-percent variation on junction uniformity, we observe that both the gain profile and quantum efficiency of Floquet TWPA are robust against up to few percent variations and only start to deteriorate rapidly after $\sigma_{Ic} > 5\%$.  $\leq5\%$ junction variation has been readily achieved and reported \cite{haygood_characterization_2019,Kreikebaum_2020}.

%

\end{document}